\documentclass[11pt]{article}
\usepackage{graphicx}
\usepackage{rotating}
\usepackage[utf8]{inputenc}
\usepackage{cmap}
\usepackage{epstopdf}
\begin{document}

 \vskip 0.5cm
  \centerline{\bf\Large Orbits of 152 Globular Clusters of the Milky Way Galaxy }
  \centerline{\bf\Large Constructed from the Gaia DR2 data}
 \bigskip
 \bigskip
  \centerline
 {
  A.~T.~ Bajkova,  V.~V. ~Bobylev
 }
 \bigskip
 \centerline{\small \it
Pulkovo Astronomical Observatory, St.-Petersburg, Russia, E-mail: bajkova@gaoran.ru}

 \bigskip
 \bigskip
 \begin{abstract}
{We present orbits and their properties for 152 globular clusters of the Milky Way galaxy obtained using average Gaia DR2 proper motions and other astrometric data from the list of Vasiliev (2019).
For orbit integrating we have used the axisymmetric model of the Galactic potential based on the Navarro-Frenk-White dark halo, and  modified by Bajkova, Bobylev (2016, 2017) using circular velocities of Galactic objects in wide region of Galactocentric distances (up to 200 kpc) from  Bhattacharjee et al. (2014) catalog. Based on the analysis of the obtained orbits, we have modified the composition of the subsystems of globular clusters presented in Massari et al. (2019).}
\end{abstract}


\section{Introduction}   
The appearance of accurate astrometric data from measurements from the Gaia satellite of the positions and spatial velocities of globular clusters (Helmi et al. 2018; Baumgardt
et al. 2019; Vasiliev 2019) makes it possible to study their dynamics, origin and evolution (Myeong et al. 2019; Massari et al. 2019; Bajkova et al. 2020).

In this work we present orbits and their properties for the almost entire list of globular clusters compiled by  Vasiliev (2019) on the basis of the most accurate measurements of their velocities and positions to date and using one of the best-fit models of the Milky Way gravitational potential.
In addition, we set a goal to revise the classification of globular clusters proposed by Massari et al. (2019) on the basis of analysis of the obtained orbits. In essence, this paper is a supplement to paper of Bajkova et al. (2020), which was devoted to the division of globular clusters into subsystems of the Galaxy, namely, bulge/bar, thick disk, and halo. Recall that in paper of Bajkova et al. (2020) a new criterion of separation of globular clusters belonging to the disk and halo of the Galaxy was proposed. This criterion is based on the bimodality of the GCs distribution $L_Z/ecc$, where $L_Z$ is $Z$ component of the angular momentum, $ecc$ is eccentricity of the orbit.

In addition to work of Bajkova et al. (2020), we present here a broader set of orbital parameters, and most importantly, we present a catalog of orbit images in two projections, which makes it possible to more fully analyze them due to visualization. Such work on visualization of the orbits of almost all globular clusters known to date, and carried out according to the latest data, has been done in the literature for the first time. This work provides, in addition to the available quantitative estimates, the possibility of a qualitative assessment of the GC dynamics, comparison of the orbits of different GCs included in different classification groups. 

In the previous work, we dealt only with the problem of dividing the GCs into formed in situ and clusters which formed in different progenitors that were only later accreted, i.e. division into disk, bulge and halo subsystems. We did not touch upon the problem of classifying the  halo GCs into subsystems, the nature of which is the accretion events of members of other galaxies (the Sausage, Sequoia) onto the Milky Way. An attempt at such a classification was made, for example, in the work of Massari et al. (2019). But we do not consider this work completed, since the analysis of the various diagrams suggests some inconsistencies and contradictions. A qualitative analysis of the orbits allowed us to make some adjustments to the Massari's classification. As a result, we offer our own modified classification, which seems to be more organic. The change in classification affected 27 objects. 

In addition to the previous article, we present the values of important orbital parameters that were not considered earlier.

This work is structured as follows. Section 2 describes the accepted most realistic model for the axially symmetric Galactic potential. Section 3 devoted to integrating the orbits and computing  parameters of the orbits. Section 4 describes data. In Section 5 we present orbits of 152 globular clusters and their properties and propose a modified classification of the GCs based on the analysis of orbits which is slightly different from the classification given by Massari et al. (2019). In Conclusions we summarize main results.

\section{Model for the Axially Symmetric Galactic Potential}

In this article the same model of gravitational potential as in work by Bajkova et al. (2020) is adopted.
The axially symmetric gravitational potential of the Galaxy is
represented as the sum of three components --- the central,
spherical bulge $\Phi_b(r(R,Z))$, the disk $\Phi_d(r(R,Z))$, and
the massive, spherical dark-matter halo $\Phi_h(r(R,Z))$:
\begin{equation}
\begin{array}{lll}
  \Phi(R,Z)=\Phi_b(r(R,Z))+\Phi_d(r(R,Z))+\Phi_h(r(R,Z)).
 \label{pot}
 \end{array}
 \end{equation}
Here, we use a cylindrical coordinate system ($R,\psi,Z$) with
its origin at the Galactic center. In Cartesian coordinates
$(X,Y,Z)$ with their origin at the Galactic center, the distance
to a star (the spherical radius) is $r^2=X^2+Y^2+Z^2=R^2+Z^2$.
The gravitational potential is expressed in units of 100 km$^2$ s$^{-2}$, distances in kpc,
masses in units of the mass of the Galaxy, $M_{gal}=2.325\times
10^7 M_\odot$, and the gravitational constant is taken to be
$G=1.$

We express the potentials of the bulge, $\Phi_b(r(R,Z))$, and
disk, $\Phi_d(r(R,Z))$, in the form suggested by Miyamoto,
Nagai (1975):
 \begin{equation}
  \Phi_b(r)=-\frac{M_b}{(r^2+b_b^2)^{1/2}},
  \label{bulge}
 \end{equation}
 \begin{equation}
 \Phi_d(R,Z)=-\frac{M_d}{\Biggl[R^2+\Bigl(a_d+\sqrt{Z^2+b_d^2}\Bigr)^2\Biggr]^{1/2}},
 \label{disk}
\end{equation}
where $M_b, M_d$ are the masses of these components, and $b_b,
a_d, b_d$ are the scale parameters of the components in kpc.

 For description of the halo component, we used the expression in Navarro-Frenk-White (NFW) form presented in Navarro et al. (1997):
 \begin{equation}
  \Phi_h(r)=-\frac{M_h}{r} \ln {\Biggl(1+\frac{r}{a_h}\Biggr)},
 \label{halo-III}
 \end{equation}
where $M_h$ is the mass, $a_h$ is the scale length.

The model of the Galactic potential, considered in this work, is the NFW model modified
in work of Bajkova \& Bobylev (2016) by fitting of the model parameters to data on HI,
maser sources and Galactic objects from Bhattacharjee et al. (2014) at distances $R$ within $\sim200$
kpc. In addition the constraints (Irrgang et al. 2013) on
the local dynamical matter density $\rho_\odot=0.1M_\odot$~pc$^{-3}$ and the force acting perpendicularly to the Galactic plane $|K_{z=1.1}|/2\pi G=77M_\odot$~pc$^{-2}$ were used.

Note that among six  models of the Galactic potential summarized in Bajkova \& Bobylev (2017) our model (denoted as Model III in papers of Bajkova \& Bobylev (2016, 2017) ensures the best fit to the data. Here we denote the model as NFWBB for short.

Parameters of the NFWBB model are given in Table 1 of work of Bajkova et al. (2020). Corresponding rotation curves up to $R=200$ kpc are shown in Fig.3 of work by Bajkova \& Bobylev (2016). When deriving the model rotation curve, we used
$R_\odot=8.3$ kpc for the Galactocentric distance of the Sun and
$V_\odot=244$ km s$^{-1}$ for the linear velocity of the Local
Standard of Rest around the center of the Galaxy. The mass of the Galaxy according to this model is $M_{G_{(R \leq 200 kpc)}}=0.75\pm0.19\times10^{12}M_\odot$. This value is consistent with the recently obtained estimate of the lower mass limit for the dark spherical NFW halo $ M_{200}=0.67^{+0.30}_{-0.15}\times10^{12}M_\odot$ (Koppelman \& Helmi, 2020) from the escape velocity using a proper motion selected halo sample.

\section{Integrating the Orbits and Computing Orbit Parameters}

The equation of motion of a test particle in an axially symmetric
gravitational potential can be obtained from the Lagrangian of the
system $\pounds$ (see Appendix A in Irrgang et al. (2013)):
\begin{equation}
 \begin{array}{lll}
 \pounds(R,Z,\dot{R},\dot{\psi},\dot{Z})=\\
 \qquad0.5(\dot{R}^2+(R\dot{\psi})^2+\dot{Z}^2)-\Phi(R,Z).
 \label{Lagr}
 \end{array}
\end{equation}
Introducing the canonical moments
\begin{equation}
 \begin{array}{lll}
    p_{R}=\partial\pounds/\partial\dot{R}=\dot{R},\\
 p_{\psi}=\partial\pounds/\partial\dot{\phi}=R^2\dot{\psi},\\
    p_{Z}=\partial\pounds/\partial\dot{Z}=\dot{Z},
 \label{moments}
 \end{array}
\end{equation}
we obtain the Lagrangian equations in the form of a system of six
first-order differential equations:
 \begin{equation}
 \begin{array}{llllll}
 \dot{R}=p_R,\\
 \dot{\psi}=p_{\psi}/R^2,\\
 \dot{Z}=p_Z,\\
 \dot{p_R}=-\partial\Phi(R,Z)/\partial R +p_{\psi}^2/R^3,\\
 \dot{p_{\psi}}=0,\\
 \dot{p_Z}=-\partial\Phi(R,Z)/\partial Z.
 \label{eq-motion}
 \end{array}
\end{equation}
We integrated Eqs. (\ref{eq-motion}) using a fourth-order Runge-Kutta algorithm.

The Sun's peculiar velocity with respect to the Local Standard of
Rest was taken to be
$(u_\odot,v_\odot,w_\odot)=(11.1,12.2,7.3)\pm(0.7,0.5,0.4)$ km s$^{-1}$
((Schonrich et al. 2010)). Here, we use the heliocentric velocities in a moving
Cartesian coordinate system with $u$ directed towards the Galactic
center, $v$ in the direction of the Galactic rotation, and $w$
perpendicular to the Galactic plane and directed towards the north
Galactic pole.

Let the initial positions and space velocities of a test particle
in the heliocentric coordinate system be
$(x_o,y_o,z_o,u_o,v_o,w_o)$. The initial positions ($X,Y,Z$) and velocities ($U,V,W$)
of the test particle in Galactic Cartesian coordinates are then
given by the formulas:
\begin{equation}
 \begin{array}{llllll}
 X=R_\odot-x_o, Y=y_o, Z=z_o+h_\odot,\\
 R=\sqrt{X^2+Y^2},\\
 U=u_o+u_\odot,\\
 V=v_o+v_\odot+V_\odot,\\
 W=w_o+w_\odot,
 \label{init}
 \end{array}
\end{equation}
where $R_\odot$ and $V_\odot$ are the Galactocentric distance and the
linear velocity of the Local Standard of Rest around the Galactic
center, $h_\odot=16$~pc (Bajkova \& Bobylev, 2016) is the height of the Sun above the Galactic plane, $\Pi$ and $\Theta$ are radial and tangential (rotational) velocities respectively.

Below we present the following orbital parameters of globular clusters:

\noindent (1)initial distance of the GC from the Galactic center $d_{GC}$:
\begin{equation}
d_{GC}=\sqrt{X^2+Y^2+Z^2};
\end{equation}

\noindent (2)radial velocity $\Pi$:
\begin{equation}
 \Pi=-U \frac{X}{R}+V \frac{Y}{R};
 \end{equation}

\noindent (3)tangential velocity $\Theta$:
\begin{equation}
 \Theta=U \frac{Y}{R}+V \frac{X}{R};
 \end{equation}

\noindent (4)total 3D velocity $V_{tot}$:
\begin{equation}
V_{tot}=\sqrt{\Pi^2+\Theta^2+W^2};
 \end{equation}

\noindent (5)apocentric distance (apo) of the orbit;

\noindent (6)pericentric distance (peri) of the orbit;

\noindent (7)the eccentricity (ecc) of the orbit:
\begin{equation}
ecc=\frac{apo-peri}{apo+peri};
 \end{equation}

\noindent (8)the components of the angular momentum:
\begin{equation}
 L_X=Y\times W-Z\times V;
 \end{equation}

 \begin{equation}
 L_Y=Z\times U-X\times W;
 \end{equation}

\begin{equation}
 L_Z=X\times V-Y\times U;
 \end{equation}

\noindent (9)inclination of the orbit $\theta$:
\begin{equation}
\theta=\arccos(\frac{L_Z}{L}),
\end{equation}
where $L=\sqrt{L_X^2+L_Y^2+L_Z^2}$;

\noindent (10)period of the orbit $T_r$;

\noindent (11)total energy $E$:
\begin{equation}
 E= \Phi(R,Z)+\frac{V_{tot}^2}{2}.
 \end{equation}

\section{Data}

In this paper, as the source of data on globular clusters the Vasiliev (2019) catalog serves. It contains average proper motions calculated from data of the Gaia DR2 Catalog, line-of-sight velocities, $(\alpha, \delta)$ positions and distances of 150 globular clusters. Data for GC Liller 1 we took from previous preprint of Vasiliev (2018). For the globular cluster FSR 1758, we took data from the work of Villanova et al. (2019).

The initial position and velocity coordinates (the 6d phase space) $(x_o,y_o,z_o,u_o,v_o,w_o)$ were calculated from these data and used for integrating the orbits. Uncertainties in the initial coordinates were calculated using Monte-Carlo simulation (1000 iterations) taking into account the measurement errors given in catalog of Vasiliev (2019). We adopted the uncertainty in the GCs heliocentric distances $d$ as 7\%, which is about 1.5 times larger than the estimate given by Vasiliev (2019).

Actually, Vasiliev (2019) used the distances from the Harris (2010) catalog, where it is assumed an error of 0.1 in distance modulus, corresponding to a relative error of 4.6\% in the distance, which, according to Vasiliev (2019), is a rather optimistic choice, since for some clusters, as follows from independent literature sources, the variation in several independent distance estimates could exceed 0.1 mag. It should be borne in mind that there is a possibility of additional error due to inhomogeneity in absorption in the Galaxy, but this is unlikely to give a contribution greater than 0.1 mag (color excess error $E (B-V)\sim0.03$). Hence, the total error (square root of the sum of squares) in the distance modulus is hardly more than 0.15 mag, which corresponds to about 7\% uncertainty in the distances.

\begin{figure*}
{\begin{center}
   \includegraphics[width=0.7\textwidth,angle=0]{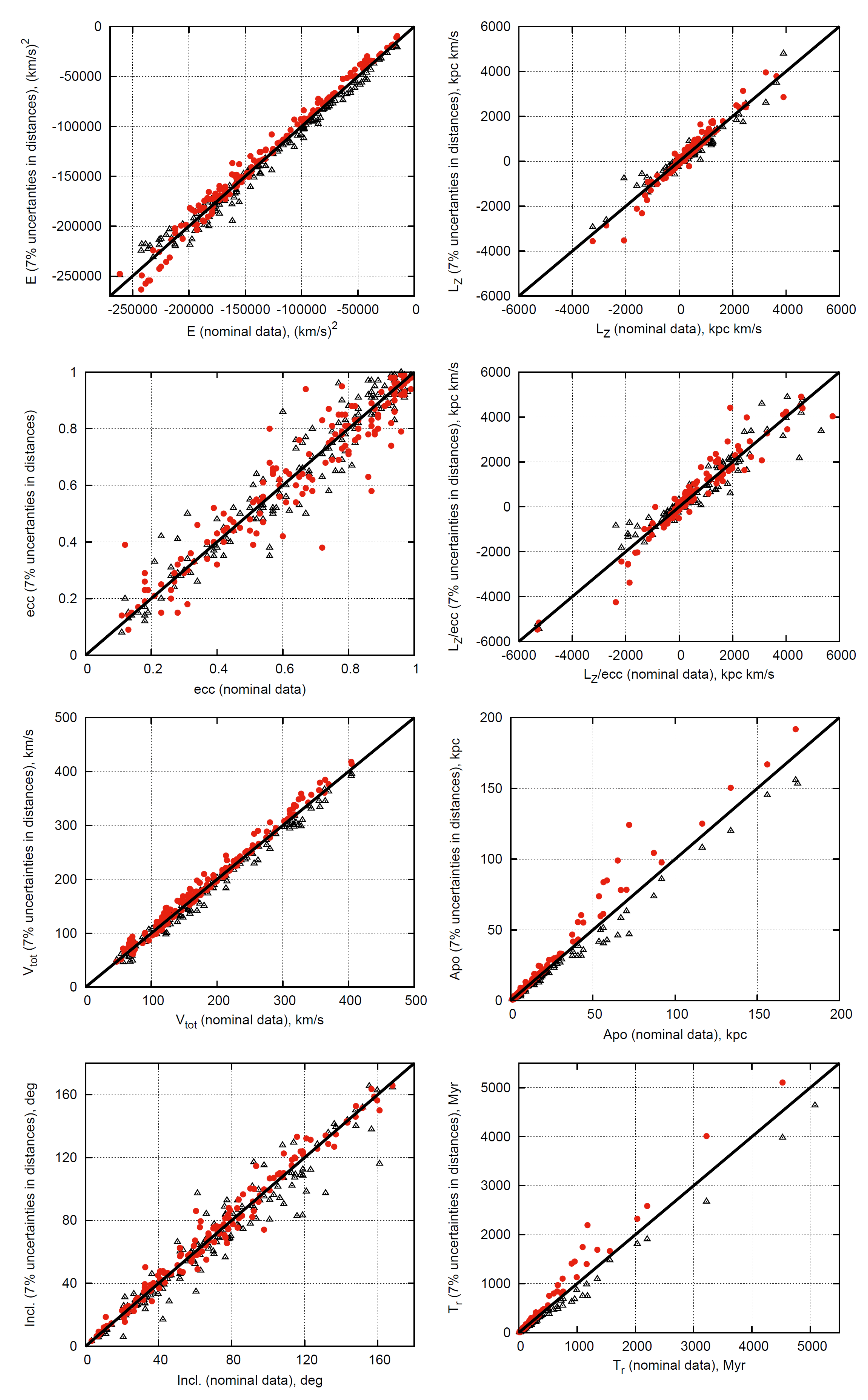}
  \caption{Comparison of the GCs orbit parameters ($E, L_Z, ecc, L_Z/ecc, V_{tot}, Apo, \theta, T_r$). The parameters obtained for the GCs with the distances which are larger on 7\% (red
points) and  less on 7\% (black triangles) than the nominal ones, are compared with the corresponding orbit parameters obtained for the GCs with nominal distances. In each
panel we plot the line of coincidence.}
\label{f_D}
\end{center}}
\end{figure*}

Even in the case of an estimate of 4.6\%, the distance seems to be the largest source of uncertainty for most of the clusters (Vasiliev (2019)), the more this is aggravated in our case, when the uncertainty in the distances is assumed to be about 1.5 times larger (7\%).
Therefore, it makes sense to study the effect of distance uncertainties on the values of the clusters orbit parameters.
Comparison of the most important orbit parameters ($E, L_Z, ecc, L_Z/ecc, V_{tot}, Apo, \theta, T_r$) is given in Fig. \ref{f_D}. In this Figure the parameters obtained for the GCs with the distances which are both larger on 7\% and  less on 7\% than the nominal ones, are compared with the corresponding orbit parameters obtained for the GCs with nominal distances.

It should be especially noted that among the analyzed parameters we also included the ratio $L_Z/ecc$, the bimodality of the distribution of which we used in work of Bajkova et al. (2020) to separate globular clusters belonging to the thick disk and halo. The influence of the uncertainty in the GC distances of 7\% led to a standard deviation of this ratio of 5\% (relative to effective range of this parameter), which we expect did not affect the result of GCs classification due to the high stability of the separation algorithm shown in work of Bajkova et al. (2020).  The greatest influence from uncertainty in distances of 7\% is experienced by such parameters as eccentricity ($ecc$) and  inclination of orbit $\theta$. The standard deviations of these parameters are 7.6 and 8.4\%, respectively. All other parameters are significantly less influenced, with a standard deviation of not larger than 3\% in each case.

\begin{table*}
 \begin{center}
  \caption[]
 {\small\baselineskip=1.0ex
Orbital properties of the GCs. For each GC
we quote values derived from orbits integrated for 5 Gyr backward.}
\bigskip
\begin{tiny}
 \label{t:prop}
 \begin{tabular}{|l|r|r|r|r|r|r|r|r|r|r|r|}\hline
 Name &$d_{GC}$&$\Pi$ &$\Theta$ &$V_{tot}$ &apo   & peri      &ecc&incl. $\theta$    &$T_r $&$L_Z$&$E$      \\
      & [kpc]  &[km/s]&[km/s]   &[km/s]    & [kpc]& [kpc]&   &[deg] &[Myr]&[kpc  &[km$^2$ \\
      &        &      &         &          &      &      &   &    &     & km/s]& /s$^2$] \\\hline
NGC 104    &   7.6  & $    6^{+  5}_{-  9}$& $  192^{+  8}_{-  4}$& $  197^{+  7}_{-  3}$& $  7.7^{+0.1}_{-0.1}$& $  5.5^{+0.3}_{-0.2}$& $0.16^{+0.02}_{-0.03}$& $  27^{+   2}_{-   1}$& $ 116 $& $   1328 $& $   -126288$\\
NGC 288    &  12.2  & $    9^{+  3}_{-  4}$& $  -42^{+ 28}_{- 22}$& $   66^{+ 16}_{- 14}$& $ 12.3^{+0.3}_{-0.4}$& $  1.4^{+0.7}_{-0.8}$& $0.80^{+0.11}_{-0.10}$& $ 119^{+   7}_{-  16}$& $ 142 $& $   -349 $& $   -116280$\\
NGC 362    &   9.5  & $  127^{+ 18}_{- 13}$& $   -1^{+ 12}_{- 12}$& $  145^{+ 16}_{-  9}$& $ 12.1^{+0.3}_{-0.6}$& $  0.1^{+0.3}_{-0.0}$& $0.99^{+0.00}_{-0.06}$& $  92^{+  22}_{-  19}$& $ 126 $& $    -10 $& $   -121290$\\
Whiting 1    &  34.7  & $ -208^{+ 18}_{- 14}$& $  109^{+ 22}_{- 16}$& $  235^{+ 18}_{- 17}$& $ 67.0^{+14.3}_{-9.0}$& $ 20.3^{+3.4}_{-1.3}$& $0.54^{+0.02}_{-0.05}$& $  67^{+   3}_{-   2}$& $1162 $& $   2494 $& $    -42265$\\
NGC 1261   &  18.2  & $  -95^{+ 11}_{-  9}$& $  -19^{+ 10}_{-  9}$& $  119^{+  9}_{-  8}$& $ 21.1^{+1.4}_{-1.0}$& $  0.7^{+0.4}_{-0.2}$& $0.94^{+0.02}_{-0.04}$& $ 121^{+  11}_{-  16}$& $ 244 $& $   -249 $& $    -91385$\\
Pal 1     &  17.4  & $   42^{+  6}_{-  8}$& $  214^{+  3}_{-  4}$& $  219^{+  3}_{-  3}$& $ 19.1^{+0.4}_{-0.9}$& $ 14.6^{+0.6}_{-0.8}$& $0.13^{+0.03}_{-0.02}$& $  14^{+   1}_{-   1}$& $ 350 $& $   3646 $& $    -78086$\\
E 1       & 124.7  & $   11^{+ 75}_{- 72}$& $ -144^{+114}_{- 88}$& $  174^{+104}_{- 45}$& $237.9^{+45.2}_{-42.9}$& $104.3^{+30.3}_{-45.9}$& $0.39^{+0.41}_{-0.00}$& $ 127^{+   6}_{-  28}$& $5594 $& $ -12100 $& $    -15914$\\
Eridanus    &  95.2  & $  -90^{+ 27}_{- 31}$& $  -28^{+ 48}_{- 27}$& $  163^{+ 43}_{- 16}$& $174.5^{+83.6}_{-24.3}$& $ 12.0^{+73.4}_{-6.2}$& $0.87^{+0.02}_{-0.31}$& $ 113^{+  13}_{-  37}$& $3220 $& $  -2064 $& $    -23988$\\
Pal 2     &  35.3  & $ -108^{+  5}_{-  5}$& $   11^{+ 22}_{- 23}$& $  108^{+  8}_{-  2}$& $ 41.0^{+2.0}_{-2.2}$& $  0.6^{+1.8}_{-0.4}$& $0.97^{+0.02}_{-0.08}$& $  42^{+  70}_{-  35}$& $ 506 $& $    373 $& $    -63896$\\
NGC 1851   &  16.9  & $  105^{+  4}_{-  4}$& $   -1^{+  5}_{-  5}$& $  133^{+  4}_{-  4}$& $ 20.1^{+0.7}_{-0.2}$& $  0.1^{+0.2}_{-0.0}$& $0.99^{+0.00}_{-0.02}$& $  92^{+   9}_{-   9}$& $ 232 $& $    -22 $& $    -94063$\\
NGC 1904   &  19.0  & $   46^{+  8}_{-  3}$& $   12^{+ 12}_{-  9}$& $   47^{+  9}_{-  0}$& $ 19.7^{+0.5}_{-0.9}$& $  0.3^{+0.5}_{-0.1}$& $0.97^{+0.01}_{-0.05}$& $  60^{+  27}_{-  25}$& $ 218 $& $    211 $& $    -96192$\\
NGC 2298   &  16.0  & $  -92^{+  6}_{-  8}$& $  -32^{+  8}_{- 10}$& $  125^{+  7}_{-  8}$& $ 18.0^{+0.7}_{-0.4}$& $  1.2^{+0.5}_{-0.4}$& $0.87^{+0.04}_{-0.05}$& $ 118^{+   8}_{-   5}$& $ 208 $& $   -500 $& $    -98716$\\
NGC 2419   &  90.2  & $   -5^{+ 10}_{- 13}$& $   47^{+ 30}_{- 17}$& $   75^{+ 31}_{- 14}$& $ 91.8^{+7.3}_{-6.1}$& $ 17.3^{+13.3}_{-5.5}$& $0.68^{+0.09}_{-0.18}$& $  52^{+  14}_{-  18}$& $1562 $& $   3907 $& $    -35896$\\
Pyxis     &  41.5  & $ -247^{+  9}_{-  5}$& $  -29^{+ 21}_{- 10}$& $  311^{+ 13}_{- 10}$& $324.2^{+95.8}_{-90.9}$& $ 41.7^{+4.7}_{-12.8}$& $0.77^{+0.07}_{-0.01}$& $  98^{+   2}_{-   6}$& $7100 $& $  -1205 $& $    -15098$\\
NGC 2808   &  11.3  & $ -157^{+  1}_{-  2}$& $   41^{+  1}_{-  7}$& $  165^{+  2}_{-  1}$& $ 14.4^{+0.7}_{-0.1}$& $  1.0^{+0.0}_{-0.1}$& $0.87^{+0.02}_{-0.00}$& $  10^{+   5}_{-   0}$& $ 158 $& $    457 $& $   -111797$\\
E 3       &   9.3  & $   44^{+ 11}_{- 16}$& $  251^{+ 13}_{- 18}$& $  276^{+ 14}_{- 18}$& $ 13.1^{+1.9}_{-2.0}$& $  9.2^{+0.4}_{-0.5}$& $0.18^{+0.06}_{-0.07}$& $  29^{+   1}_{-   1}$& $ 224 $& $   2240 $& $    -97684$\\
Pal 3     &  95.9  & $ -147^{+ 47}_{- 70}$& $   89^{+ 79}_{- 62}$& $  184^{+ 81}_{- 20}$& $173.4^{+77.4}_{-73.9}$& $ 76.3^{+29.1}_{-13.7}$& $0.39^{+0.38}_{-0.00}$& $  67^{+  15}_{-  14}$& $4524 $& $   6495 $& $    -20091$\\
NGC 3201   &   9.1  & $ -114^{+ 17}_{- 18}$& $ -301^{+  9}_{-  9}$& $  355^{+ 12}_{- 11}$& $ 26.3^{+3.7}_{-3.0}$& $  8.4^{+0.3}_{-0.3}$& $0.52^{+0.05}_{-0.05}$& $ 152^{+   1}_{-   1}$& $ 372 $& $  -2728 $& $    -75483$\\
Pal 4     & 111.4  & $  -25^{+ 46}_{- 67}$& $  -33^{+ 63}_{- 60}$& $   70^{+ 72}_{-  0}$& $116.5^{+22.1}_{-8.5}$& $ 16.3^{+60.0}_{-1.6}$& $0.75^{+0.01}_{-0.49}$& $ 103^{+   7}_{-  17}$& $2032 $& $  -1394 $& $    -31065$\\
Crater    & 144.8  & $ -101^{+110}_{- 98}$& $  -63^{+105}_{-270}$& $  135^{+249}_{-  0}$& $156.1^{+97.5}_{-90.0}$& $117.4^{+42.6}_{-18.0}$& $0.14^{+0.73}_{-0.00}$& $ 108^{+  25}_{-  26}$& $5082 $& $  -6104 $& $    -18859$\\
NGC 4147   &  21.5  & $   47^{+  9}_{-  8}$& $   -3^{+ 12}_{- 22}$& $  136^{+  5}_{-  5}$& $ 26.4^{+0.6}_{-1.9}$& $  0.4^{+0.9}_{-0.0}$& $0.97^{+0.00}_{-0.07}$& $  93^{+  20}_{-  12}$& $ 314 $& $    -27 $& $    -81117$\\
NGC 4372   &   7.3  & $   16^{+ 11}_{- 14}$& $  133^{+  6}_{-  9}$& $  150^{+  6}_{-  7}$& $  7.3^{+0.3}_{-0.2}$& $  3.0^{+0.2}_{-0.3}$& $0.42^{+0.04}_{-0.02}$& $  28^{+   3}_{-   2}$& $  98 $& $    962 $& $   -139570$\\
Rup 106    &  18.5  & $ -242^{+  5}_{-  8}$& $   91^{+ 16}_{- 15}$& $  261^{+ 12}_{-  7}$& $ 37.9^{+5.6}_{-5.1}$& $  4.7^{+0.9}_{-0.8}$& $0.78^{+0.03}_{-0.03}$& $  46^{+   6}_{-   4}$& $ 498 $& $   1640 $& $    -64986$\\
NGC 4590   &  10.3  & $ -169^{+  7}_{- 12}$& $  293^{+  6}_{- 12}$& $  339^{+  4}_{-  6}$& $ 29.9^{+1.4}_{-2.4}$& $  8.9^{+0.2}_{-0.4}$& $0.54^{+0.02}_{-0.02}$& $  41^{+   2}_{-   3}$& $ 428 $& $   2453 $& $    -70495$\\
NGC 4833   &   7.2  & $  105^{+ 12}_{- 15}$& $   40^{+ 12}_{- 13}$& $  120^{+ 12}_{- 13}$& $  8.0^{+0.4}_{-0.3}$& $  0.7^{+0.2}_{-0.3}$& $0.83^{+0.07}_{-0.04}$& $  36^{+  13}_{-   8}$& $  86 $& $    286 $& $   -144428$\\
NGC 5024   &  18.5  & $  -95^{+  7}_{-  6}$& $  141^{+  8}_{- 10}$& $  184^{+  8}_{- 10}$& $ 22.3^{+2.1}_{-2.0}$& $  8.9^{+1.4}_{-1.4}$& $0.43^{+0.03}_{-0.03}$& $  74^{+   2}_{-   2}$& $ 332 $& $    797 $& $    -80241$\\
NGC 5053   &  17.9  & $  -89^{+  5}_{-  3}$& $  134^{+  5}_{-  9}$& $  164^{+  4}_{-  8}$& $ 18.0^{+1.1}_{-0.6}$& $ 10.4^{+1.0}_{-1.1}$& $0.27^{+0.04}_{-0.03}$& $  76^{+   1}_{-   1}$& $ 300 $& $    727 $& $    -85221$\\
NGC 5139   &   6.6  & $  -70^{+  7}_{-  4}$& $  -72^{+  6}_{-  7}$& $  128^{+  7}_{-  9}$& $  7.4^{+0.1}_{-0.4}$& $  1.1^{+0.3}_{-0.1}$& $0.73^{+0.02}_{-0.05}$& $ 137^{+   7}_{-   5}$& $  80 $& $   -462 $& $   -147850$\\
NGC 5272   &  12.2  & $  -38^{+  7}_{-  5}$& $  143^{+ 10}_{-  7}$& $  200^{+  8}_{-  6}$& $ 15.9^{+0.7}_{-0.9}$& $  5.2^{+0.4}_{-0.3}$& $0.51^{+0.01}_{-0.03}$& $  57^{+   1}_{-   2}$& $ 212 $& $    994 $& $    -98226$\\
NGC 5286   &   8.9  & $ -220^{+  4}_{-  4}$& $  -44^{+ 14}_{- 15}$& $  224^{+  3}_{-  1}$& $ 13.7^{+1.2}_{-0.7}$& $  0.8^{+0.4}_{-0.2}$& $0.89^{+0.02}_{-0.04}$& $ 123^{+  10}_{-  10}$& $ 154 $& $   -375 $& $   -113332$\\
NGC 5466   &  16.4  & $  172^{+ 17}_{- 22}$& $ -141^{+ 29}_{- 17}$& $  317^{+  8}_{- 17}$& $ 53.7^{+5.7}_{-10.4}$& $  5.9^{+0.6}_{-1.1}$& $0.80^{+0.02}_{-0.02}$& $ 108^{+   2}_{-   2}$& $ 750 $& $   -820 $& $    -52693$\\
NGC 5634   &  21.1  & $  -45^{+ 12}_{- 16}$& $   39^{+ 15}_{- 16}$& $   65^{+ 11}_{-  5}$& $ 21.6^{+1.1}_{-1.1}$& $  2.1^{+0.9}_{-0.3}$& $0.82^{+0.02}_{-0.06}$& $  70^{+   7}_{-   3}$& $ 256 $& $    346 $& $    -89143$\\
NGC 5694   &  29.3  & $ -185^{+  8}_{- 13}$& $  -44^{+ 18}_{- 22}$& $  252^{+ 12}_{- 10}$& $ 70.4^{+10.4}_{-7.8}$& $  2.5^{+1.7}_{-1.0}$& $0.93^{+0.03}_{-0.04}$& $ 136^{+  10}_{-  17}$& $ 992 $& $  -1019 $& $    -45290$\\
IC 4499   &  15.6  & $ -245^{+  7}_{-  4}$& $  -75^{+  9}_{- 14}$& $  263^{+  4}_{-  5}$& $ 30.0^{+3.1}_{-2.7}$& $  6.5^{+0.8}_{-0.4}$& $0.65^{+0.02}_{-0.05}$& $ 113^{+   4}_{-   3}$& $ 406 $& $  -1066 $& $    -72235$\\
NGC 5824   &  25.7  & $  -41^{+ 17}_{- 20}$& $  105^{+ 13}_{- 24}$& $  215^{+ 11}_{- 14}$& $ 37.4^{+2.6}_{-4.3}$& $ 14.0^{+1.7}_{-3.6}$& $0.45^{+0.10}_{-0.05}$& $  58^{+   4}_{-   2}$& $ 598 $& $   2393 $& $    -59779$\\
Pal 5     &  18.4  & $  -54^{+  5}_{-  6}$& $  160^{+ 39}_{- 43}$& $  169^{+ 37}_{- 38}$& $ 18.9^{+2.4}_{-0.6}$& $ 10.7^{+4.3}_{-3.6}$& $0.28^{+0.18}_{-0.13}$& $  66^{+   4}_{-   4}$& $ 308 $& $   1260 $& $    -83194$\\
NGC 5897   &   7.3  & $   88^{+ 13}_{- 26}$& $   97^{+ 14}_{- 23}$& $  159^{+  8}_{- 19}$& $  8.7^{+0.6}_{-0.4}$& $  1.9^{+0.4}_{-0.5}$& $0.64^{+0.08}_{-0.05}$& $  61^{+   5}_{-   5}$& $ 106 $& $    362 $& $   -131771$\\
NGC 5904   &   6.3  & $ -290^{+ 17}_{- 15}$& $  126^{+ 11}_{- 14}$& $  365^{+ 13}_{- 14}$& $ 23.3^{+3.9}_{-3.3}$& $  2.3^{+0.5}_{-0.3}$& $0.82^{+0.03}_{-0.03}$& $  72^{+   3}_{-   3}$& $ 286 $& $    402 $& $    -85576$\\
NGC 5927   &   4.7  & $  -39^{+ 20}_{- 23}$& $  233^{+  6}_{- 11}$& $  236^{+  6}_{-  9}$& $  5.2^{+0.3}_{-0.2}$& $  4.2^{+0.4}_{-0.5}$& $0.11^{+0.07}_{-0.05}$& $   9^{+   1}_{-   1}$& $  82 $& $   1077 $& $   -148662$\\
NGC 5946   &   5.8  & $   35^{+  6}_{- 18}$& $   25^{+  6}_{- 15}$& $  116^{+  9}_{-  9}$& $  5.9^{+0.6}_{-0.2}$& $  0.4^{+0.2}_{-0.2}$& $0.89^{+0.05}_{-0.05}$& $  77^{+   8}_{-   3}$& $  66 $& $    141 $& $   -158006$\\
ESO 224-8   &  12.6  & $  -43^{+ 60}_{- 14}$& $  259^{+ 21}_{- 22}$& $  262^{+ 19}_{- 21}$& $ 17.0^{+3.0}_{-2.7}$& $ 11.9^{+0.9}_{-1.3}$& $0.18^{+0.08}_{-0.07}$& $   7^{+   1}_{-   1}$& $ 292 $& $   3245 $& $    -85205$\\
NGC 5986   &   4.7  & $   62^{+ 21}_{- 33}$& $   23^{+ 16}_{- 15}$& $   68^{+ 24}_{- 32}$& $  5.2^{+0.6}_{-0.0}$& $  0.2^{+0.2}_{-0.1}$& $0.93^{+0.05}_{-0.08}$& $  66^{+  10}_{-  14}$& $  56 $& $     94 $& $   -168505$\\
FSR 1716   &   4.8  & $   87^{+ 47}_{- 44}$& $  228^{+ 21}_{- 15}$& $  281^{+ 24}_{- 10}$& $  7.0^{+1.4}_{-0.7}$& $  3.9^{+0.7}_{-0.7}$& $0.28^{+0.13}_{-0.07}$& $  31^{+   2}_{-   3}$& $ 106 $& $   1089 $& $   -137191$\\
Pal 14    &  71.4  & $  117^{+ 25}_{- 18}$& $   16^{+ 16}_{- 41}$& $  177^{+ 29}_{- 17}$& $133.9^{+31.3}_{-22.5}$& $  2.2^{+5.2}_{-0.4}$& $0.97^{+0.00}_{-0.07}$& $  50^{+  63}_{-   0}$& $2202 $& $    794 $& $    -29458$\\
BH 184    &   4.4  & $   40^{+ 15}_{- 21}$& $  121^{+  9}_{- 11}$& $  156^{+  8}_{-  9}$& $  4.7^{+0.4}_{-0.4}$& $  1.7^{+0.2}_{-0.3}$& $0.47^{+0.06}_{-0.04}$& $  36^{+   3}_{-   3}$& $  58 $& $    531 $& $   -168579$\\
NGC 6093   &   3.7  & $   33^{+ 12}_{- 17}$& $   16^{+  9}_{- 19}$& $   71^{+  9}_{-  7}$& $  4.3^{+0.0}_{-0.9}$& $  0.2^{+0.3}_{-0.1}$& $0.93^{+0.05}_{-0.13}$& $  83^{+   8}_{-   5}$& $  44 $& $     25 $& $   -176933$\\
NGC 6121   &   6.3  & $  -52^{+  2}_{-  2}$& $   10^{+ 18}_{- 12}$& $   54^{+  7}_{-  2}$& $  6.4^{+0.2}_{-0.1}$& $  0.2^{+0.1}_{-0.1}$& $0.94^{+0.04}_{-0.03}$& $  21^{+  82}_{-  31}$& $  68 $& $     60 $& $   -159021$\\
NGC 6101   &  11.1  & $  -12^{+ 29}_{- 12}$& $ -314^{+  3}_{-  2}$& $  370^{+  6}_{-  4}$& $ 44.2^{+4.1}_{-4.8}$& $ 10.9^{+0.4}_{-0.9}$& $0.61^{+0.02}_{-0.02}$& $ 143^{+   2}_{-   2}$& $ 658 $& $  -3236 $& $    -56722$\\
NGC 6144   &   2.7  & $  -69^{+ 90}_{- 93}$& $ -196^{+ 66}_{- 15}$& $  213^{+  7}_{-  3}$& $  3.2^{+0.6}_{-0.0}$& $  2.1^{+0.2}_{-0.3}$& $0.21^{+0.14}_{-0.01}$& $ 114^{+   8}_{-   9}$& $  40 $& $   -239 $& $   -172662$\\
NGC 6139   &   3.5  & $   -1^{+ 16}_{- 16}$& $   76^{+  4}_{-  8}$& $  156^{+  9}_{-  9}$& $  3.6^{+0.3}_{-0.1}$& $  1.1^{+0.2}_{-0.2}$& $0.54^{+0.07}_{-0.03}$& $  62^{+   4}_{-   3}$& $  52 $& $    248 $& $   -176480$\\
Terzan 3   &   2.5  & $  -61^{+ 45}_{- 51}$& $  206^{+  7}_{- 22}$& $  236^{+  7}_{-  8}$& $  3.2^{+0.4}_{-0.2}$& $  2.2^{+0.1}_{-0.4}$& $0.18^{+0.13}_{-0.03}$& $  42^{+   6}_{-   4}$& $  44 $& $    440 $& $   -175324$\\
\hline
 \end{tabular}
 \end{tiny}
  \end{center}
  \end{table*}

\begin{table*}
 \begin{center}
\centerline{\small Table~1: continued.}
\bigskip
\begin{tiny}
 \label{t:prop1}
 \begin{tabular}{|l|r|r|r|r|r|r|r|r|r|r|r|}\hline
 Name &$d_{GC}$&$\Pi$ &$\Theta$ &$V_{tot}$ &apo   & peri      &ecc&incl. $\theta$    &$T_r $&$L_Z$&$E$      \\
      & [kpc]  &[km/s]&[km/s]   &[km/s]    & [kpc]& [kpc]&   &[deg] &[Myr]&[kpc  &[km$^2$ \\
      &        &      &         &          &      &      &   &    &     & km/s]& /s$^2$] \\\hline
NGC 6171   &   3.5  & $   -4^{+  7}_{-  3}$& $   78^{+ 15}_{-  8}$& $  101^{+ 13}_{-  6}$& $  3.8^{+0.3}_{-0.3}$& $  0.6^{+0.3}_{-0.2}$& $0.72^{+0.08}_{-0.11}$& $  52^{+   7}_{-   4}$& $  44 $& $    191 $& $   -178959$\\
ESO 452-11   &   2.1  & $  -24^{+ 11}_{- 15}$& $  -13^{+ 12}_{- 11}$& $  107^{+ 10}_{-  2}$& $  2.9^{+0.4}_{-0.1}$& $  0.0^{+0.1}_{-0.0}$& $0.97^{+0.01}_{-0.03}$& $ 101^{+   7}_{-  10}$& $  26 $& $    -16 $& $   -202684$\\
NGC 6205   &   8.6  & $   20^{+  6}_{-  4}$& $  -26^{+  5}_{-  7}$& $   87^{+  5}_{-  6}$& $  8.6^{+0.3}_{-0.2}$& $  1.0^{+0.2}_{-0.1}$& $0.79^{+0.03}_{-0.03}$& $ 105^{+   4}_{-   3}$& $ 100 $& $   -187 $& $   -134416$\\
NGC 6229   &  29.9  & $   30^{+ 11}_{- 12}$& $    6^{+  8}_{-  9}$& $   59^{+  8}_{- 13}$& $ 31.0^{+1.1}_{-1.0}$& $  0.5^{+0.6}_{-0.1}$& $0.97^{+0.01}_{-0.04}$& $  77^{+  26}_{-  21}$& $ 376 $& $    135 $& $    -74434$\\
NGC 6218   &   4.8  & $   -9^{+  3}_{-  9}$& $  135^{+  6}_{- 10}$& $  158^{+  6}_{-  7}$& $  5.0^{+0.4}_{-0.1}$& $  2.3^{+0.3}_{-0.2}$& $0.37^{+0.05}_{-0.02}$& $  37^{+   2}_{-   2}$& $  64 $& $    581 $& $   -158135$\\
FSR 1735   &   4.3  & $ -102^{+ 17}_{-  7}$& $   -5^{+ 15}_{- 21}$& $  225^{+ 16}_{- 12}$& $  5.3^{+0.6}_{-0.3}$& $  1.0^{+0.5}_{-0.1}$& $0.69^{+0.03}_{-0.10}$& $  92^{+   6}_{-   4}$& $  68 $& $    -22 $& $   -157538$\\
NGC 6235   &   4.0  & $  159^{+  8}_{-  5}$& $  197^{+ 28}_{- 40}$& $  256^{+ 24}_{- 30}$& $  6.2^{+1.2}_{-1.1}$& $  2.7^{+0.4}_{-0.5}$& $0.39^{+0.05}_{-0.03}$& $  53^{+  11}_{-   7}$& $  84 $& $    570 $& $   -145439$\\
NGC 6254   &   4.8  & $  -87^{+  4}_{-  7}$& $  134^{+ 13}_{- 16}$& $  167^{+  9}_{- 10}$& $  5.2^{+0.2}_{-0.3}$& $  2.1^{+0.3}_{-0.4}$& $0.42^{+0.06}_{-0.06}$& $  36^{+   4}_{-   1}$& $  78 $& $    606 $& $   -157472$\\
NGC 6256   &   2.9  & $ -170^{+ 18}_{-  0}$& $   28^{+ 39}_{- 18}$& $  198^{+  4}_{- 11}$& $  4.4^{+0.9}_{-0.8}$& $  0.1^{+0.3}_{-0.1}$& $0.94^{+0.04}_{-0.15}$& $  78^{+   7}_{-  14}$& $  46 $& $     79 $& $   -181873$\\
Pal 15    &  38.2  & $  154^{+  9}_{- 13}$& $   -5^{+ 26}_{- 17}$& $  162^{+  9}_{- 11}$& $ 54.6^{+2.8}_{-4.4}$& $  1.2^{+3.0}_{-0.3}$& $0.96^{+0.01}_{-0.11}$& $  98^{+  18}_{-  35}$& $ 726 $& $   -165 $& $    -53233$\\
NGC 6266   &   2.0  & $   42^{+ 17}_{- 15}$& $  122^{+ 10}_{- 18}$& $  146^{+  6}_{- 11}$& $  2.5^{+0.5}_{-0.4}$& $  0.6^{+0.3}_{-0.1}$& $0.62^{+0.04}_{-0.10}$& $  32^{+   4}_{-   4}$& $  32 $& $    215 $& $   -205537$\\
NGC 6273   &   1.6  & $  -98^{+ 67}_{-143}$& $ -240^{+218}_{- 34}$& $  315^{+  8}_{-  4}$& $  3.8^{+0.6}_{-0.3}$& $  1.0^{+0.3}_{-0.1}$& $0.59^{+0.11}_{-0.08}$& $ 109^{+  10}_{-  18}$& $  48 $& $   -144 $& $   -172941$\\
NGC 6284   &   7.3  & $   14^{+  2}_{-  2}$& $   -2^{+  7}_{- 19}$& $  113^{+ 12}_{-  6}$& $  7.5^{+0.9}_{-0.9}$& $  0.7^{+0.3}_{-0.2}$& $0.83^{+0.04}_{-0.06}$& $  91^{+  10}_{-   4}$& $  90 $& $    -16 $& $   -142332$\\
NGC 6287   &   2.0  & $ -301^{+264}_{-107}$& $  -64^{+ 40}_{- 80}$& $  318^{+  5}_{-  4}$& $  5.3^{+0.6}_{-0.5}$& $  0.8^{+0.1}_{-0.1}$& $0.75^{+0.04}_{-0.02}$& $  95^{+   3}_{-   3}$& $  66 $& $    -60 $& $   -159009$\\
NGC 6293   &   1.8  & $ -152^{+ 52}_{- 33}$& $  -80^{+115}_{- 36}$& $  233^{+  7}_{- 10}$& $  3.6^{+0.4}_{-1.0}$& $  0.2^{+0.2}_{-0.1}$& $0.91^{+0.08}_{-0.17}$& $ 131^{+   7}_{-  39}$& $  38 $& $    -93 $& $   -191358$\\
NGC 6304   &   2.5  & $   79^{+  5}_{- 10}$& $  191^{+  7}_{-  6}$& $  219^{+  5}_{-  6}$& $  3.3^{+0.4}_{-0.6}$& $  1.8^{+0.2}_{-0.4}$& $0.29^{+0.04}_{-0.02}$& $  20^{+   2}_{-   2}$& $  52 $& $    474 $& $   -183132$\\
NGC 6316   &   2.4  & $  103^{+  9}_{-  7}$& $   51^{+ 26}_{- 33}$& $  143^{+ 16}_{- 11}$& $  3.0^{+0.5}_{-0.6}$& $  0.4^{+0.1}_{-0.3}$& $0.77^{+0.18}_{-0.03}$& $  40^{+  31}_{-  16}$& $  36 $& $    106 $& $   -197316$\\
NGC 6341   &   9.8  & $   53^{+  6}_{-  4}$& $   13^{+  9}_{-  5}$& $  109^{+ 12}_{- 13}$& $ 10.7^{+0.3}_{-0.3}$& $  0.3^{+0.2}_{-0.1}$& $0.94^{+0.02}_{-0.04}$& $  79^{+   4}_{-   9}$& $ 124 $& $    108 $& $   -125312$\\
NGC 6325   &   1.3  & $  -81^{+ 29}_{- 48}$& $ -181^{+187}_{- 63}$& $  214^{+ 17}_{- 19}$& $  1.3^{+0.4}_{-0.1}$& $  1.1^{+0.2}_{-0.3}$& $0.12^{+0.16}_{-0.00}$& $ 114^{+  13}_{-  25}$& $  18 $& $   -107 $& $   -212097$\\
NGC 6333   &   1.8  & $  -89^{+122}_{-118}$& $  346^{+  6}_{- 48}$& $  364^{+  5}_{-  3}$& $  6.4^{+0.8}_{-0.2}$& $  1.0^{+0.2}_{-0.1}$& $0.74^{+0.02}_{-0.03}$& $  59^{+   2}_{-   6}$& $  74 $& $    327 $& $   -151409$\\
NGC 6342   &   1.6  & $  -25^{+ 70}_{- 89}$& $  164^{+ 13}_{- 66}$& $  168^{+  6}_{-  4}$& $  1.7^{+0.7}_{-0.3}$& $  0.9^{+0.2}_{-0.4}$& $0.31^{+0.29}_{-0.12}$& $  64^{+   6}_{-   3}$& $  24 $& $    117 $& $   -207010$\\
NGC 6356   &   7.2  & $   47^{+  5}_{-  9}$& $  107^{+ 25}_{- 12}$& $  160^{+ 17}_{-  7}$& $  7.9^{+1.3}_{-1.1}$& $  2.5^{+1.0}_{-0.5}$& $0.52^{+0.06}_{-0.10}$& $  43^{+   3}_{-   4}$& $ 104 $& $    713 $& $   -136303$\\
NGC 6355   &   1.2  & $ -207^{+ 70}_{- 47}$& $ -110^{+108}_{- 54}$& $  275^{+  8}_{- 10}$& $  2.2^{+1.1}_{-0.5}$& $  0.6^{+0.2}_{-0.2}$& $0.56^{+0.17}_{-0.10}$& $ 106^{+   8}_{-  10}$& $  28 $& $    -95 $& $   -199192$\\
NGC 6352   &   3.6  & $   42^{+ 18}_{- 13}$& $  226^{+  5}_{- 13}$& $  230^{+  4}_{- 10}$& $  4.1^{+0.6}_{-0.3}$& $  3.2^{+0.2}_{-0.3}$& $0.13^{+0.05}_{-0.03}$& $  12^{+   1}_{-   1}$& $  68 $& $    794 $& $   -163864$\\
IC 1257   &  17.6  & $  -45^{+  7}_{- 12}$& $  -50^{+ 12}_{- 18}$& $   70^{+ 12}_{-  0}$& $ 18.1^{+1.1}_{-0.8}$& $  1.8^{+0.8}_{-0.4}$& $0.82^{+0.04}_{-0.08}$& $ 158^{+   4}_{-  11}$& $ 208 $& $   -817 $& $    -98556$\\
Terzan 2   &   1.0  & $ -120^{+ 62}_{- 23}$& $  -47^{+ 13}_{- 48}$& $  136^{+  3}_{-  2}$& $  1.2^{+0.3}_{-0.4}$& $  0.1^{+0.1}_{-0.0}$& $0.86^{+0.05}_{-0.19}$& $ 161^{+   3}_{-  19}$& $  14 $& $    -44 $& $   -242360$\\
NGC 6366   &   5.3  & $   94^{+  3}_{-  3}$& $  134^{+  2}_{-  6}$& $  175^{+  2}_{-  5}$& $  5.8^{+0.2}_{-0.2}$& $  2.2^{+0.1}_{-0.2}$& $0.45^{+0.03}_{-0.00}$& $  32^{+   2}_{-   1}$& $  74 $& $    699 $& $   -153378$\\
Terzan 4   &   1.2  & $   15^{+  8}_{- 39}$& $   75^{+ 18}_{- 13}$& $  124^{+ 14}_{-  8}$& $  1.3^{+0.2}_{-0.4}$& $  0.2^{+0.1}_{-0.1}$& $0.68^{+0.03}_{-0.10}$& $  52^{+   7}_{-   6}$& $  14 $& $     92 $& $   -234494$\\
BH 229    &   0.5  & $    7^{+ 46}_{- 39}$& $  -55^{+ 48}_{-  0}$& $  292^{+ 14}_{- 22}$& $  0.8^{+1.1}_{-0.1}$& $  0.3^{+0.4}_{-0.2}$& $0.49^{+0.32}_{-0.18}$& $ 100^{+   0}_{-   9}$& $  10 $& $    -21 $& $   -249808$\\
FSR 1758   &   3.7  & $   60^{+ 30}_{- 60}$& $ -347^{+  8}_{-  6}$& $  405^{+  8}_{-  8}$& $ 14.3^{+2.3}_{-2.1}$& $  3.7^{+0.4}_{-0.5}$& $0.59^{+0.03}_{-0.02}$& $ 148^{+   2}_{-   2}$& $ 178 $& $  -1275 $& $   -106508$\\
NGC 6362   &   5.2  & $   17^{+ 14}_{- 28}$& $  124^{+ 11}_{-  6}$& $  160^{+ 11}_{-  4}$& $  5.3^{+0.2}_{-0.4}$& $  2.5^{+0.2}_{-0.3}$& $0.37^{+0.04}_{-0.04}$& $  45^{+   2}_{-   4}$& $  70 $& $    583 $& $   -153468$\\
Liller 1   &   0.8  & $  107^{+ 25}_{- 57}$& $  -56^{+ 61}_{- 43}$& $  123^{+ 28}_{- 31}$& $  0.8^{+0.2}_{-0.1}$& $  0.1^{+0.2}_{-0.1}$& $0.81^{+0.13}_{-0.16}$& $ 155^{+  16}_{-  74}$& $   8 $& $    -42 $& $   -261395$\\
NGC6380   &   3.1  & $  -62^{+  5}_{- 14}$& $  -35^{+ 16}_{-  9}$& $   72^{+ 14}_{-  7}$& $  3.4^{+0.2}_{-0.6}$& $  0.2^{+0.1}_{-0.1}$& $0.89^{+0.05}_{-0.07}$& $ 168^{+  11}_{-  26}$& $  40 $& $   -105 $& $   -194646$\\
Terzan 1   &   1.6  & $  -73^{+  3}_{-  5}$& $   63^{+ 10}_{- 20}$& $   96^{+  8}_{- 11}$& $  1.8^{+0.5}_{-0.6}$& $  0.2^{+0.1}_{-0.1}$& $0.79^{+0.06}_{-0.04}$& $  11^{+  10}_{-   6}$& $  22 $& $    102 $& $   -224803$\\
Pismis 26   &   1.4  & $ -112^{+111}_{- 58}$& $  204^{+ 32}_{- 55}$& $  307^{+  9}_{-  8}$& $  3.2^{+0.8}_{-0.1}$& $  0.9^{+0.7}_{-0.4}$& $0.56^{+0.17}_{-0.20}$& $  41^{+   9}_{-   2}$& $  44 $& $    271 $& $   -186876$\\
NGC 6388   &   3.0  & $  -66^{+ 24}_{- 19}$& $  -94^{+ 18}_{- 12}$& $  116^{+  6}_{-  7}$& $  3.5^{+0.4}_{-0.2}$& $  0.7^{+0.4}_{-0.2}$& $0.69^{+0.07}_{-0.13}$& $ 148^{+   8}_{-   9}$& $  46 $& $   -257 $& $   -190150$\\
NGC 6402   &   4.0  & $  -20^{+ 23}_{- 21}$& $   48^{+ 10}_{-  7}$& $   57^{+ 16}_{-  7}$& $  4.8^{+0.2}_{-0.5}$& $  0.3^{+0.2}_{-0.0}$& $0.88^{+0.00}_{-0.08}$& $  47^{+   6}_{-  10}$& $  52 $& $    158 $& $   -176457$\\
NGC 6401   &   2.5  & $  -30^{+ 27}_{- 18}$& $ -254^{+  4}_{-  4}$& $  302^{+  8}_{-  4}$& $  4.5^{+0.8}_{-0.7}$& $  2.4^{+0.6}_{-0.6}$& $0.31^{+0.07}_{-0.03}$& $ 143^{+   2}_{-   3}$& $  62 $& $   -595 $& $   -161761$\\
NGC 6397   &   6.3  & $   35^{+  7}_{-  6}$& $  127^{+ 11}_{-  6}$& $  179^{+  9}_{-  4}$& $  6.5^{+0.1}_{-0.2}$& $  2.8^{+0.3}_{-0.2}$& $0.40^{+0.03}_{-0.04}$& $  43^{+   2}_{-   4}$& $  86 $& $    796 $& $   -144538$\\
Pal 6     &   2.5  & $ -191^{+  2}_{-  3}$& $   21^{+ 16}_{- 14}$& $  246^{+  6}_{-  8}$& $  4.5^{+0.4}_{-0.8}$& $  0.1^{+0.1}_{-0.0}$& $0.96^{+0.01}_{-0.03}$& $  83^{+   4}_{-   6}$& $  44 $& $     52 $& $   -179351$\\
NGC 6426   &  14.3  & $ -112^{+ 27}_{- 16}$& $   93^{+  7}_{- 20}$& $  148^{+ 11}_{- 25}$& $ 16.6^{+1.1}_{-0.7}$& $  3.2^{+0.5}_{-0.9}$& $0.67^{+0.08}_{-0.03}$& $  27^{+   4}_{-   4}$& $ 202 $& $   1216 $& $   -100666$\\
Djorg 1    &   1.2  & $ -252^{+271}_{-102}$& $  315^{+ 49}_{- 51}$& $  404^{+ 12}_{-  9}$& $  5.9^{+1.6}_{-1.6}$& $  0.8^{+0.1}_{-0.2}$& $0.76^{+0.05}_{-0.06}$& $  21^{+  13}_{-   7}$& $  66 $& $    351 $& $   -161437$\\
Terzan 5   &   1.5  & $   84^{+ 11}_{-  6}$& $   70^{+ 13}_{- 23}$& $  114^{+  9}_{-  7}$& $  1.7^{+0.5}_{-0.6}$& $  0.2^{+0.1}_{-0.1}$& $0.78^{+0.04}_{-0.05}$& $  33^{+  18}_{-   7}$& $  20 $& $    104 $& $   -226313$\\
NGC 6440   &   1.3  & $   91^{+ 18}_{- 36}$& $  -42^{+ 54}_{- 34}$& $  107^{+ 10}_{-  8}$& $  1.4^{+0.4}_{-0.0}$& $  0.2^{+0.1}_{-0.1}$& $0.78^{+0.15}_{-0.07}$& $ 116^{+  21}_{-  34}$& $  14 $& $    -49 $& $   -231796$\\
NGC 6441   &   3.6  & $   16^{+ 15}_{-  8}$& $   66^{+ 18}_{- 19}$& $   71^{+ 20}_{- 16}$& $  3.6^{+0.7}_{-0.6}$& $  0.8^{+0.2}_{-0.2}$& $0.66^{+0.09}_{-0.07}$& $  21^{+  11}_{-   6}$& $  42 $& $    228 $& $   -186312$\\
Terzan 6   &   1.5  & $ -138^{+  7}_{-  3}$& $  -51^{+ 21}_{- 32}$& $  153^{+  9}_{-  3}$& $  2.0^{+0.2}_{-0.6}$& $  0.1^{+0.1}_{-0.1}$& $0.86^{+0.05}_{-0.11}$& $ 157^{+  17}_{-  26}$& $  22 $& $    -77 $& $   -220185$\\
NGC 6453   &   3.4  & $ -105^{+  9}_{-  6}$& $   38^{+ 30}_{- 15}$& $  194^{+ 19}_{-  5}$& $  3.9^{+0.8}_{-0.6}$& $  0.9^{+0.5}_{-0.1}$& $0.61^{+0.04}_{-0.14}$& $  78^{+   5}_{-   8}$& $  54 $& $    129 $& $   -172906$\\
NGC 6496   &   4.0  & $  -37^{+ 76}_{- 63}$& $  320^{+ 24}_{- 41}$& $  328^{+ 26}_{- 28}$& $  9.1^{+2.1}_{-1.6}$& $  3.7^{+0.4}_{-0.6}$& $0.42^{+0.10}_{-0.07}$& $  32^{+   5}_{-   3}$& $ 120 $& $   1111 $& $   -126509$\\
Terzan 9   &   1.3  & $  -50^{+  8}_{-  7}$& $   22^{+ 18}_{- 14}$& $   76^{+  7}_{-  2}$& $  1.4^{+0.6}_{-0.4}$& $  0.1^{+0.1}_{-0.0}$& $0.92^{+0.03}_{-0.08}$& $  70^{+  12}_{-  13}$& $  16 $& $     29 $& $   -236057$\\
Djorg 2    &   2.0  & $  161^{+  9}_{-  6}$& $  155^{+  6}_{- 11}$& $  228^{+  3}_{-  4}$& $  3.2^{+0.6}_{-0.7}$& $  0.9^{+0.3}_{-0.3}$& $0.57^{+0.06}_{-0.04}$& $  11^{+   1}_{-   1}$& $  42 $& $    316 $& $   -192751$\\
NGC 6517   &   4.0  & $   55^{+  8}_{- 17}$& $   33^{+ 10}_{- 12}$& $   73^{+  5}_{- 11}$& $  4.6^{+0.4}_{-0.4}$& $  0.2^{+0.1}_{-0.1}$& $0.91^{+0.03}_{-0.04}$& $  58^{+  10}_{-  10}$& $  50 $& $    127 $& $   -179281$\\
Terzan 10   &   2.2  & $  231^{+ 20}_{- 52}$& $   87^{+ 64}_{- 22}$& $  343^{+ 20}_{- 13}$& $  5.9^{+1.4}_{-1.8}$& $  0.7^{+0.3}_{-0.1}$& $0.79^{+0.04}_{-0.11}$& $  72^{+   4}_{-  10}$& $  76 $& $    193 $& $   -155422$\\
\hline
 \end{tabular}
 \end{tiny}
  \end{center}
  \end{table*}

\begin{table*}
 \begin{center}
\centerline{\small Table~1: continued.}
\bigskip
\begin{tiny}
 \label{t:prop2}
 \begin{tabular}{|l|r|r|r|r|r|r|r|r|r|r|r|}\hline
 Name &$d_{GC}$&$\Pi$ &$\Theta$ &$V_{tot}$ &apo   & peri      &ecc&incl. $\theta$    &$T_r $&$L_Z$&$E$      \\
      & [kpc]  &[km/s]&[km/s]   &[km/s]    & [kpc]& [kpc]&   &[deg] &[Myr]&[kpc  &[km$^2$ \\
      &        &      &         &          &      &      &   &    &     & km/s]& /s$^2$] \\\hline
NGC 6522   &   0.8  & $   34^{+ 25}_{- 15}$& $   92^{+ 23}_{- 57}$& $  213^{+ 17}_{-  9}$& $  1.2^{+0.5}_{-0.3}$& $  0.2^{+0.3}_{-0.2}$& $0.67^{+0.14}_{-0.15}$& $  63^{+  16}_{-   6}$& $  16 $& $     58 $& $   -238519$\\
NGC 6535   &   4.0  & $   93^{+  9}_{- 10}$& $  -83^{+ 13}_{-  6}$& $  133^{+  2}_{-  6}$& $  4.6^{+0.4}_{-0.3}$& $  1.0^{+0.1}_{-0.2}$& $0.64^{+0.06}_{-0.03}$& $ 160^{+   1}_{-   3}$& $  56 $& $   -320 $& $   -173024$\\
NGC 6528   &   0.7  & $ -197^{+300}_{- 67}$& $  113^{+ 31}_{- 67}$& $  229^{+  2}_{-  2}$& $  1.0^{+0.9}_{-0.2}$& $  0.3^{+0.0}_{-0.2}$& $0.60^{+0.33}_{-0.14}$& $  70^{+   7}_{-   5}$& $  14 $& $     51 $& $   -241883$\\
NGC 6539   &   3.1  & $    1^{+ 28}_{-  9}$& $  118^{+  5}_{-  3}$& $  208^{+ 14}_{-  9}$& $  3.4^{+0.3}_{-0.2}$& $  1.9^{+0.2}_{-0.3}$& $0.30^{+0.08}_{-0.03}$& $  56^{+   1}_{-   2}$& $  56 $& $    347 $& $   -174387$\\
NGC 6540   &   3.0  & $   13^{+  3}_{-  2}$& $  148^{+  9}_{-  7}$& $  159^{+  8}_{-  5}$& $  3.1^{+0.4}_{-0.4}$& $  1.6^{+0.3}_{-0.2}$& $0.32^{+0.04}_{-0.04}$& $  22^{+   2}_{-   1}$& $  48 $& $    448 $& $   -187517$\\
NGC 6544   &   5.3  & $    6^{+  2}_{-  2}$& $    6^{+ 14}_{- 10}$& $   91^{+  8}_{-  7}$& $  5.6^{+0.2}_{-0.1}$& $  0.1^{+0.2}_{-0.0}$& $0.98^{+0.00}_{-0.06}$& $  86^{+   7}_{-   9}$& $  52 $& $     31 $& $   -166630$\\
NGC 6541   &   2.3  & $  123^{+ 33}_{- 62}$& $  192^{+ 25}_{- 20}$& $  254^{+ 11}_{-  6}$& $  3.8^{+0.6}_{-0.6}$& $  1.3^{+0.2}_{-0.1}$& $0.50^{+0.08}_{-0.10}$& $  40^{+   6}_{-   4}$& $  50 $& $    334 $& $   -174968$\\
ESO 280-06   &  13.8  & $   31^{+  5}_{- 10}$& $   16^{+ 10}_{- 25}$& $   91^{+  8}_{- 11}$& $ 14.2^{+1.1}_{-1.3}$& $  0.9^{+0.5}_{-0.4}$& $0.88^{+0.05}_{-0.05}$& $  77^{+  20}_{-   7}$& $ 164 $& $    210 $& $   -109791$\\
NGC 6553   &   2.4  & $   45^{+ 15}_{- 12}$& $  245^{+  2}_{-  4}$& $  250^{+  2}_{-  3}$& $  3.3^{+0.3}_{-0.4}$& $  2.3^{+0.4}_{-0.5}$& $0.19^{+0.04}_{-0.04}$& $   7^{+   2}_{-   1}$& $  52 $& $    588 $& $   -179839$\\
NGC 6558   &   1.2  & $  187^{+ 60}_{- 94}$& $   93^{+ 36}_{- 60}$& $  209^{+  4}_{-  4}$& $  1.7^{+0.4}_{-0.6}$& $  0.3^{+0.2}_{-0.0}$& $0.72^{+0.08}_{-0.29}$& $  63^{+  15}_{-   8}$& $  20 $& $     87 $& $   -217130$\\
Pal 7     &   3.9  & $  -74^{+ 12}_{-  9}$& $  270^{+  7}_{-  4}$& $  281^{+  7}_{-  4}$& $  6.0^{+0.6}_{-0.4}$& $  3.5^{+0.3}_{-0.1}$& $0.26^{+0.02}_{-0.02}$& $  11^{+   1}_{-   1}$& $  86 $& $   1042 $& $   -147032$\\
Terzan 12   &   3.6  & $  -94^{+  5}_{-  4}$& $  172^{+  8}_{-  9}$& $  219^{+  6}_{-  6}$& $  4.4^{+0.4}_{-0.5}$& $  2.2^{+0.2}_{-0.3}$& $0.33^{+0.04}_{-0.03}$& $  28^{+   2}_{-   2}$& $  62 $& $    625 $& $   -167749$\\
NGC 6569   &   2.8  & $  -40^{+  2}_{-  4}$& $  174^{+ 18}_{- 26}$& $  180^{+ 18}_{- 25}$& $  3.0^{+0.6}_{-0.7}$& $  1.9^{+0.3}_{-0.7}$& $0.23^{+0.17}_{-0.05}$& $  26^{+  12}_{-   5}$& $  56 $& $    440 $& $   -182361$\\
ESO 456-78   &   2.0  & $   71^{+ 10}_{- 10}$& $  199^{+  5}_{-  4}$& $  252^{+  7}_{-  6}$& $  2.9^{+0.6}_{-0.2}$& $  1.4^{+0.5}_{-0.3}$& $0.34^{+0.06}_{-0.07}$& $  34^{+   2}_{-   2}$& $  52 $& $    373 $& $   -186632$\\
NGC 6584   &   6.8  & $  197^{+  7}_{- 30}$& $   98^{+ 22}_{- 34}$& $  324^{+ 11}_{- 22}$& $ 18.0^{+2.1}_{-2.8}$& $  1.7^{+0.7}_{-0.6}$& $0.83^{+0.05}_{-0.06}$& $  52^{+  12}_{-   3}$& $ 212 $& $    556 $& $    -98089$\\
NGC 6624   &   1.2  & $  -29^{+ 48}_{- 24}$& $   60^{+ 14}_{- 18}$& $  136^{+  7}_{-  4}$& $  1.5^{+0.3}_{-0.0}$& $  0.2^{+0.1}_{-0.2}$& $0.78^{+0.18}_{-0.05}$& $  73^{+   3}_{-   6}$& $  20 $& $     37 $& $   -226410$\\
NGC 6626   &   3.0  & $  -28^{+  2}_{-  4}$& $   57^{+  9}_{- 12}$& $  113^{+  8}_{-  5}$& $  3.1^{+0.5}_{-0.1}$& $  0.5^{+0.1}_{-0.2}$& $0.75^{+0.10}_{-0.04}$& $  60^{+   6}_{-   4}$& $  42 $& $    169 $& $   -193067$\\
NGC 6638   &   2.0  & $   68^{+ 16}_{- 12}$& $   14^{+ 13}_{- 21}$& $   74^{+ 14}_{-  7}$& $  2.4^{+0.5}_{-0.1}$& $  0.1^{+0.0}_{-0.0}$& $0.95^{+0.03}_{-0.03}$& $  80^{+  16}_{-   9}$& $  20 $& $     22 $& $   -212068$\\
NGC 6637   &   1.6  & $   35^{+ 27}_{-111}$& $   91^{+  6}_{- 81}$& $  129^{+  8}_{-  8}$& $  2.2^{+0.2}_{-0.4}$& $  0.2^{+0.0}_{-0.1}$& $0.87^{+0.11}_{-0.01}$& $  77^{+  17}_{-   5}$& $  22 $& $     40 $& $   -212500$\\
NGC 6642   &   1.7  & $  112^{+  6}_{- 22}$& $   25^{+ 32}_{- 51}$& $  125^{+  8}_{- 11}$& $  2.2^{+0.3}_{-0.1}$& $  0.1^{+0.2}_{-0.0}$& $0.94^{+0.02}_{-0.11}$& $  46^{+  78}_{-  17}$& $  24 $& $     36 $& $   -215457$\\
NGC 6652   &   2.5  & $  -54^{+  6}_{-  4}$& $   28^{+  7}_{- 18}$& $  186^{+ 11}_{- 16}$& $  4.2^{+0.2}_{-0.3}$& $  0.1^{+0.1}_{-0.1}$& $0.96^{+0.03}_{-0.05}$& $  76^{+   9}_{-   5}$& $  38 $& $     42 $& $   -183595$\\
NGC 6656   &   5.2  & $  176^{+  3}_{-  2}$& $  201^{+  2}_{-  2}$& $  303^{+  6}_{-  5}$& $  9.8^{+0.7}_{-0.5}$& $  3.1^{+0.2}_{-0.1}$& $0.53^{+0.01}_{-0.01}$& $  33^{+   2}_{-   2}$& $ 126 $& $   1044 $& $   -125471$\\
Pal 8     &   5.3  & $  -21^{+ 13}_{- 25}$& $  117^{+ 23}_{- 15}$& $  122^{+ 27}_{- 15}$& $  5.6^{+0.3}_{-0.6}$& $  1.8^{+0.6}_{-0.4}$& $0.51^{+0.06}_{-0.11}$& $  23^{+   3}_{-   2}$& $  72 $& $    593 $& $   -160020$\\
NGC 6681   &   2.0  & $  221^{+ 34}_{-166}$& $   55^{+135}_{- 44}$& $  287^{+  9}_{-  3}$& $  4.5^{+0.3}_{-0.4}$& $  0.7^{+0.5}_{-0.1}$& $0.74^{+0.10}_{-0.15}$& $  84^{+   8}_{-   9}$& $  52 $& $     36 $& $   -167768$\\
NGC 6712   &   3.6  & $  146^{+  6}_{-  7}$& $   26^{+  9}_{- 17}$& $  208^{+  7}_{- 10}$& $  5.5^{+0.3}_{-0.2}$& $  0.2^{+0.1}_{-0.1}$& $0.93^{+0.04}_{-0.02}$& $  78^{+   7}_{-   4}$& $  56 $& $     93 $& $   -168544$\\
NGC 6715   &  18.6  & $  232^{+  6}_{-  8}$& $   49^{+ 25}_{- 22}$& $  317^{+ 21}_{- 11}$& $ 56.4^{+24.2}_{-10.1}$& $ 14.8^{+1.8}_{-1.0}$& $0.58^{+0.07}_{-0.03}$& $  81^{+   4}_{-   4}$& $ 906 $& $    850 $& $    -48159$\\
NGC 6717   &   2.5  & $  -10^{+ 31}_{- 13}$& $  116^{+  8}_{-  9}$& $  119^{+  9}_{-  7}$& $  2.8^{+0.4}_{-0.4}$& $  0.7^{+0.2}_{-0.1}$& $0.59^{+0.06}_{-0.09}$& $  32^{+   7}_{-   4}$& $  36 $& $    251 $& $   -195762$\\
NGC 6723   &   2.6  & $  100^{+  6}_{-192}$& $ -178^{+353}_{-  3}$& $  208^{+  5}_{-  5}$& $  3.1^{+0.4}_{-0.0}$& $  1.8^{+0.2}_{-0.0}$& $0.26^{+0.05}_{-0.03}$& $  90^{+   8}_{-  12}$& $  40 $& $     -2 $& $   -175356$\\
NGC 6749   &   5.0  & $  -23^{+ 17}_{- 10}$& $  110^{+ 10}_{-  9}$& $  112^{+ 10}_{-  8}$& $  5.1^{+0.4}_{-0.2}$& $  1.6^{+0.2}_{-0.1}$& $0.53^{+0.05}_{-0.05}$& $   3^{+   1}_{-   0}$& $  62 $& $    556 $& $   -167068$\\
NGC 6752   &   5.5  & $  -23^{+  7}_{-  2}$& $  179^{+  3}_{-  6}$& $  190^{+  3}_{-  6}$& $  5.7^{+0.3}_{-0.3}$& $  3.6^{+0.2}_{-0.2}$& $0.23^{+0.02}_{-0.02}$& $  24^{+   1}_{-   0}$& $  82 $& $    931 $& $   -147164$\\
NGC 6760   &   5.0  & $   92^{+ 10}_{-  9}$& $  147^{+  7}_{-  8}$& $  174^{+  7}_{-  7}$& $  5.6^{+0.3}_{-0.3}$& $  2.2^{+0.2}_{-0.2}$& $0.44^{+0.04}_{-0.03}$& $   6^{+   1}_{-   0}$& $  72 $& $    724 $& $   -158732$\\
NGC 6779   &   9.3  & $  155^{+  3}_{-  2}$& $  -15^{+  6}_{-  9}$& $  185^{+  6}_{-  4}$& $ 12.4^{+0.4}_{-0.6}$& $  0.3^{+0.2}_{-0.1}$& $0.96^{+0.01}_{-0.03}$& $ 101^{+   7}_{-   5}$& $ 134 $& $   -135 $& $   -119696$\\
Terzan 7   &  15.3  & $  260^{+  6}_{-  6}$& $   25^{+ 17}_{- 11}$& $  320^{+ 11}_{- 11}$& $ 42.9^{+7.7}_{-8.3}$& $ 12.8^{+0.6}_{-1.1}$& $0.54^{+0.05}_{-0.05}$& $  86^{+   2}_{-   3}$& $ 668 $& $    335 $& $    -56545$\\
Pal 10    &   6.6  & $  -56^{+ 15}_{-  5}$& $  186^{+  8}_{- 13}$& $  195^{+  7}_{- 14}$& $  7.0^{+0.3}_{-0.3}$& $  4.0^{+0.4}_{-0.4}$& $0.27^{+0.04}_{-0.03}$& $   8^{+   1}_{-   1}$& $ 100 $& $   1234 $& $   -138495$\\
Arp 2     &  21.2  & $  243^{+  7}_{-  8}$& $   68^{+ 20}_{- 11}$& $  311^{+ 10}_{- 15}$& $ 65.1^{+21.7}_{-13.5}$& $ 17.8^{+1.7}_{-1.6}$& $0.57^{+0.06}_{-0.04}$& $  78^{+   2}_{-   4}$& $1094 $& $   1256 $& $    -43628$\\
NGC 6809   &   4.1  & $ -199^{+  5}_{-  7}$& $   76^{+ 15}_{- 19}$& $  220^{+  5}_{-  3}$& $  5.7^{+0.4}_{-0.5}$& $  1.2^{+0.3}_{-0.3}$& $0.66^{+0.06}_{-0.07}$& $  67^{+   6}_{-   3}$& $  76 $& $    266 $& $   -154417$\\
Terzan 8   &  19.1  & $  269^{+  8}_{-  5}$& $   37^{+ 16}_{- 23}$& $  315^{+ 12}_{-  7}$& $ 58.5^{+13.5}_{-6.8}$& $ 16.0^{+0.8}_{-0.8}$& $0.57^{+0.06}_{-0.03}$& $  84^{+   4}_{-   3}$& $ 958 $& $    584 $& $    -46785$\\
Pal 11    &   8.1  & $  -16^{+ 28}_{- 20}$& $  139^{+ 12}_{- 10}$& $  140^{+ 14}_{-  9}$& $  8.2^{+0.7}_{-0.4}$& $  3.5^{+0.6}_{-0.4}$& $0.40^{+0.05}_{-0.04}$& $  27^{+   3}_{-   2}$& $ 108 $& $   1013 $& $   -132057$\\
NGC 6838   &   7.0  & $   39^{+  5}_{- 11}$& $  204^{+  5}_{-  4}$& $  212^{+  4}_{-  4}$& $  7.3^{+0.2}_{-0.2}$& $  5.0^{+0.3}_{-0.2}$& $0.18^{+0.02}_{-0.03}$& $  12^{+   1}_{-   1}$& $ 112 $& $   1423 $& $   -132307$\\
NGC 6864   &  14.6  & $  -99^{+ 14}_{-  6}$& $   18^{+ 10}_{- 25}$& $  111^{+  7}_{- 14}$& $ 16.4^{+1.3}_{-0.4}$& $  0.5^{+0.6}_{-0.5}$& $0.94^{+0.06}_{-0.05}$& $  61^{+  41}_{-   9}$& $ 186 $& $    209 $& $   -103615$\\
NGC 6934   &  12.7  & $ -289^{+ 20}_{- 18}$& $  103^{+ 20}_{- 30}$& $  330^{+ 15}_{- 17}$& $ 40.9^{+7.8}_{-5.8}$& $  2.5^{+0.6}_{-0.8}$& $0.88^{+0.04}_{-0.03}$& $  23^{+   5}_{-   2}$& $ 520 $& $   1204 $& $    -63341$\\
NGC 6981   &  12.8  & $ -154^{+ 10}_{- 13}$& $   -4^{+ 14}_{- 25}$& $  230^{+ 10}_{-  7}$& $ 22.3^{+2.1}_{-1.3}$& $  0.3^{+0.3}_{-0.1}$& $0.97^{+0.01}_{-0.03}$& $ 108^{+  25}_{-  36}$& $ 252 $& $    -35 $& $    -89722$\\
NGC 7006   &  38.5  & $ -140^{+  6}_{-  8}$& $  -33^{+ 17}_{-  7}$& $  168^{+  9}_{- 12}$& $ 56.3^{+2.3}_{-4.5}$& $  2.9^{+0.7}_{-1.5}$& $0.90^{+0.05}_{-0.02}$& $ 133^{+  13}_{-  13}$& $ 760 $& $  -1170 $& $    -52073$\\
NGC 7078   &  10.6  & $    8^{+ 11}_{- 12}$& $  118^{+  8}_{- 11}$& $  122^{+  9}_{- 10}$& $ 10.6^{+0.5}_{-0.5}$& $  3.5^{+0.4}_{-0.5}$& $0.50^{+0.05}_{-0.04}$& $  28^{+   3}_{-   1}$& $ 140 $& $   1119 $& $   -119947$\\
NGC 7089   &  10.4  & $  170^{+  5}_{-  6}$& $  -18^{+ 12}_{- 15}$& $  243^{+  8}_{-  8}$& $ 18.8^{+1.7}_{-1.0}$& $  0.6^{+0.2}_{-0.3}$& $0.94^{+0.03}_{-0.02}$& $ 119^{+  12}_{-  20}$& $ 214 $& $   -147 $& $    -97640$\\
NGC 7099   &   7.2  & $  -32^{+ 14}_{-  9}$& $  -55^{+ 16}_{- 12}$& $  126^{+  8}_{-  9}$& $  8.2^{+0.6}_{-0.5}$& $  1.0^{+0.3}_{-0.3}$& $0.78^{+0.06}_{-0.05}$& $ 119^{+   3}_{-   6}$& $  94 $& $   -234 $& $   -137480$\\
Pal 12    &  15.7  & $  146^{+ 22}_{- 37}$& $  304^{+ 22}_{- 24}$& $  356^{+ 12}_{- 17}$& $ 72.0^{+18.9}_{-16.0}$& $ 15.5^{+1.0}_{-0.9}$& $0.65^{+0.05}_{-0.08}$& $  67^{+   2}_{-   2}$& $1178 $& $   2142 $& $    -41826$\\
Pal 13    &  27.0  & $  268^{+  6}_{- 11}$& $  -78^{+ 21}_{- 17}$& $  289^{+  9}_{- 12}$& $ 87.1^{+8.6}_{-13.4}$& $  8.3^{+0.9}_{-1.9}$& $0.83^{+0.03}_{-0.02}$& $ 115^{+   8}_{-   6}$& $1346 $& $  -1586 $& $    -38661$\\
NGC 7492   &  25.4  & $  -87^{+ 16}_{- 22}$& $  -13^{+  9}_{-  7}$& $  108^{+ 19}_{- 10}$& $ 28.2^{+2.5}_{-2.3}$& $  3.1^{+1.4}_{-1.0}$& $0.80^{+0.06}_{-0.07}$& $  95^{+   6}_{-   5}$& $ 352 $& $   -120 $& $    -77041$\\
\hline
 \end{tabular}
 \end{tiny}
  \end{center}
  \end{table*}

\section{Orbits of 152 Globular Clusters and their Properties}
\begin{figure*}
{\begin{center}
   \includegraphics[width=0.92\textwidth,angle=0]{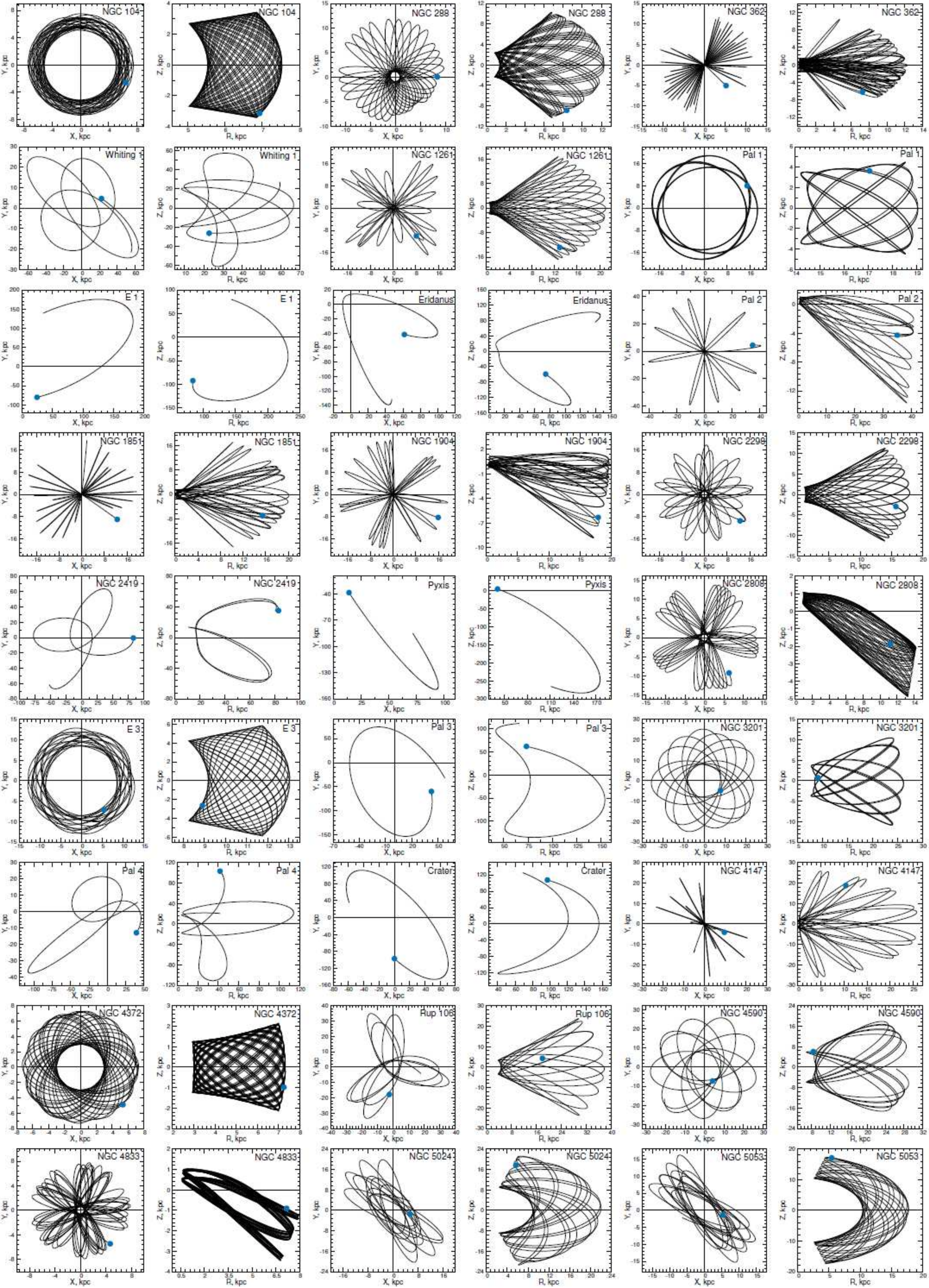}
  \caption{Orbits of the GCs obtained by integrating for 5 Gyr backward. (X,Y) and (R,Z) projections are given. The blue filled circle indicates the beginning of the orbit.}
\label{f1}
\end{center}}
\end{figure*}

The orbits of the 152 globular clusters were obtained by integrating Eg. (\ref{eq-motion}) for 5 Gyr backward. The (X,Y) and (R,Z) orbit projections for each of all GCs are presented in Fig. \ref{f1}.
The orbit properties are presented in Table \ref{t:prop}. To calculate the uncertainties in the orbit properties, we used the Monte Carlo method with 100 realizations, taking into account the uncertainties in the initial coordinates and velocities of GCs, as well as errors in the peculiar velocity of the Sun. For parameters $T_r, L_Z$ and $E$, we give in Table \ref{t:prop} only nominal values to save space.

\begin{table*}
 \begin{center}
 \caption[]
 {\small\baselineskip=1.0ex
 Globular clusters with modified classification.
  }
  \bigskip
 \begin{small}
 \label{t:crd}
 \begin{tabular}{|l|r|r||l|r|r||l|r|r|}\hline
 Name & GS &GS(m)& Name&GS&GS(m)&Name&GS &GS(m) \\\hline
E 3         &  H99  & D & NGC 6254    &  LE   & D  & NGC 6553     &   B  & D  \\
NGC 5466    &  Seq  & GE& NGC 6304    &  B    & D  & NGC 6569     &   B  & D  \\
NGC 5634    &  H99  &GE & Liller 1    &  XXX  & B  & ESO 456-78   &   B  & D  \\
NGC 5694    &  HE   & GE& NGC 6388    &   B   & Seq& NGC 6584     &   HE & GE \\
NGC 5824    &  Sgr  &H99& NGC 6401    &   LE  & Seq& NGC 6712     &   LE & GE \\
NGC 5904    &  H99  & GE& NGC 6426    &   HE  & H99& NGC 6934     &   HE & GE \\
Pal 14      &  HE   & GE& NGC 6539    &    B  & D  & NGC 6981     &   H99& GE \\
NGC 6144    &  LE   &Seq& NGC 6540    &    B  & D  & NGC 7006     &   Seq& GE \\
NGC 6235    &  GE   & D & NGC 6544    &   LE  & GE & Pal 13       &   Seq& GE \\
\hline
 \end{tabular}
 \end{small}
  \end{center}
  \end{table*}

Based on the analysis of orbits and their properties, a small regrouping of globular clusters by subsystems has been made. Table \ref{t:crd} shows 27 GCs, which have changed their belonging to one group or another. In this table, the column designated as GS -- Galactic Subsystem, gives the classification proposed by Massari et al. (2019), and the column designated as GS(m), gives a modified classification. The following designations for the Galactic subsystems are used here: D (disk), B (bulge), GE (Gaia-Enceladus, or Gaia-Sausage), H99 (Helmi Streams), Seq (Sequoia galaxy), Sgr (Sagittarius dwarf), HE (unassociated High-Energy), LE (unassociated Low-Energy), XXX (clusters with no available kinematics).

The separation of globular clusters into subsystems of the bulge/bar, thick disk, and halo was performed by us in work Bajkova et al. (2020) using a criterion based on the bimodal distribution of globular clusters over parameter $L_Z/ecc$. The composition of the bulge/bar and thick disk reflects these results. In the redistribution of the remaining globular clusters between the halo subsystems (GE, Seq, Sgr, H99), we took into account the parameters of the orbits. For example, globular clusters with strong radially elongated orbits were assigned to GE. The rest of the rearrangement also took into account the proximity of the orbital shapes.

In Fig. \ref{fElz}  the "$L_Z $-- Energy", "$L_Z/ecc$-- Energy" and "Radial velocity -- Rotational velocity" diagrams are presented. These diagrams are given both for classification of GCs by Massari et al. (2019) and for the modified one for comparison. As you can see from Fig. \ref{f1}, the modified classification looks like a more correct one from the point of view of a greater similarity of the orbits of globular clusters included in their subsystem. From a comparison of the diagrams "$L_Z$ -- Energy", "$L_Z/ecc$ -- Energy" and "Radial velocity -- Rotational velocity", it can also be concluded that the modified classification is more organic.

\begin{figure*}
{\begin{center}
   \includegraphics[width=0.92\textwidth,angle=0]{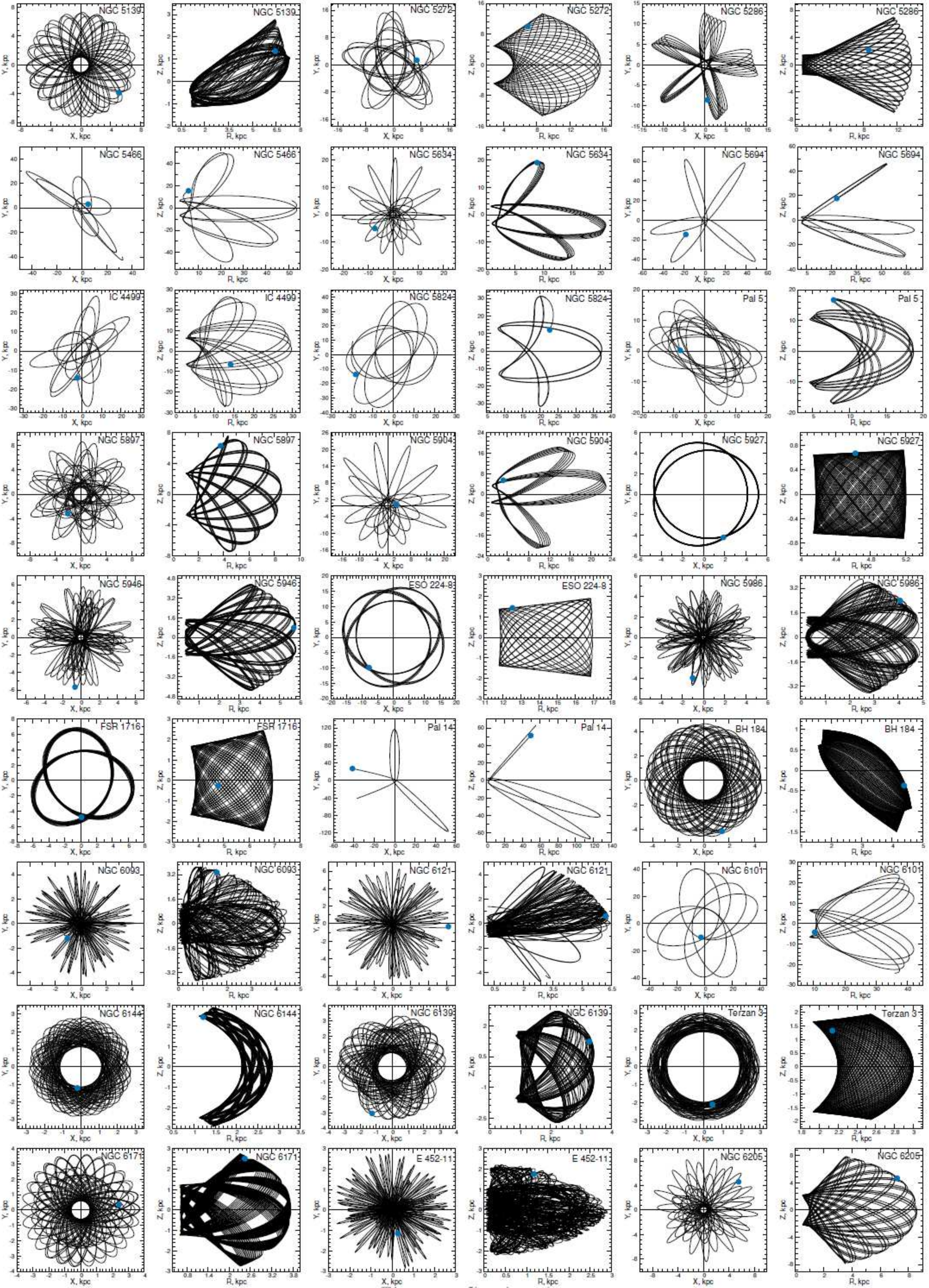}
 \centerline{Fig.~2 - continued}
\label{f2}
\end{center}}
\end{figure*}

\begin{figure*}
{\begin{center}
   \includegraphics[width=0.92\textwidth,angle=0]{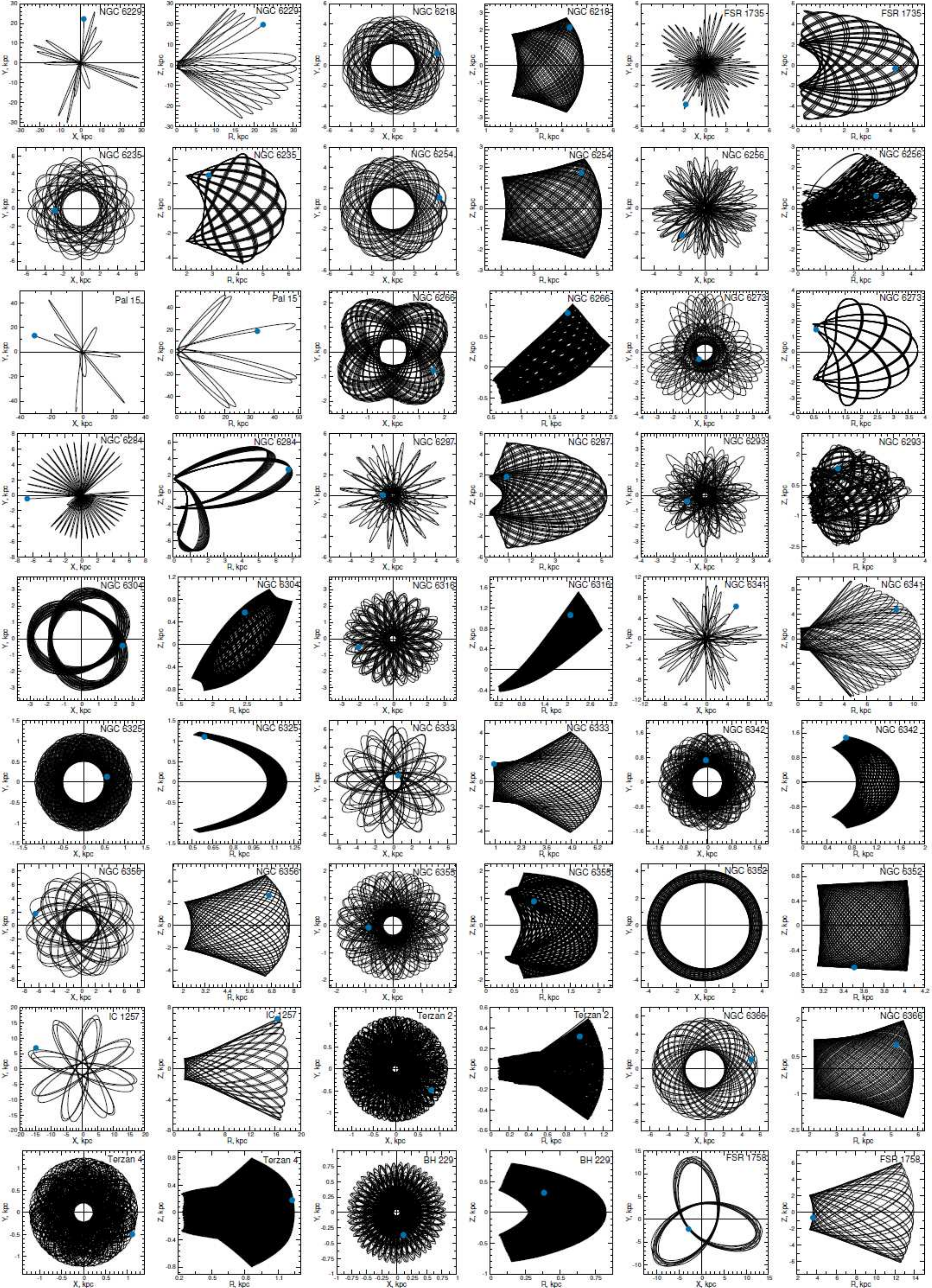}
 \centerline{Fig.~2 - continued}
\label{f3}
\end{center}}
\end{figure*}

\begin{figure*}
{\begin{center}
   \includegraphics[width=0.92\textwidth,angle=0]{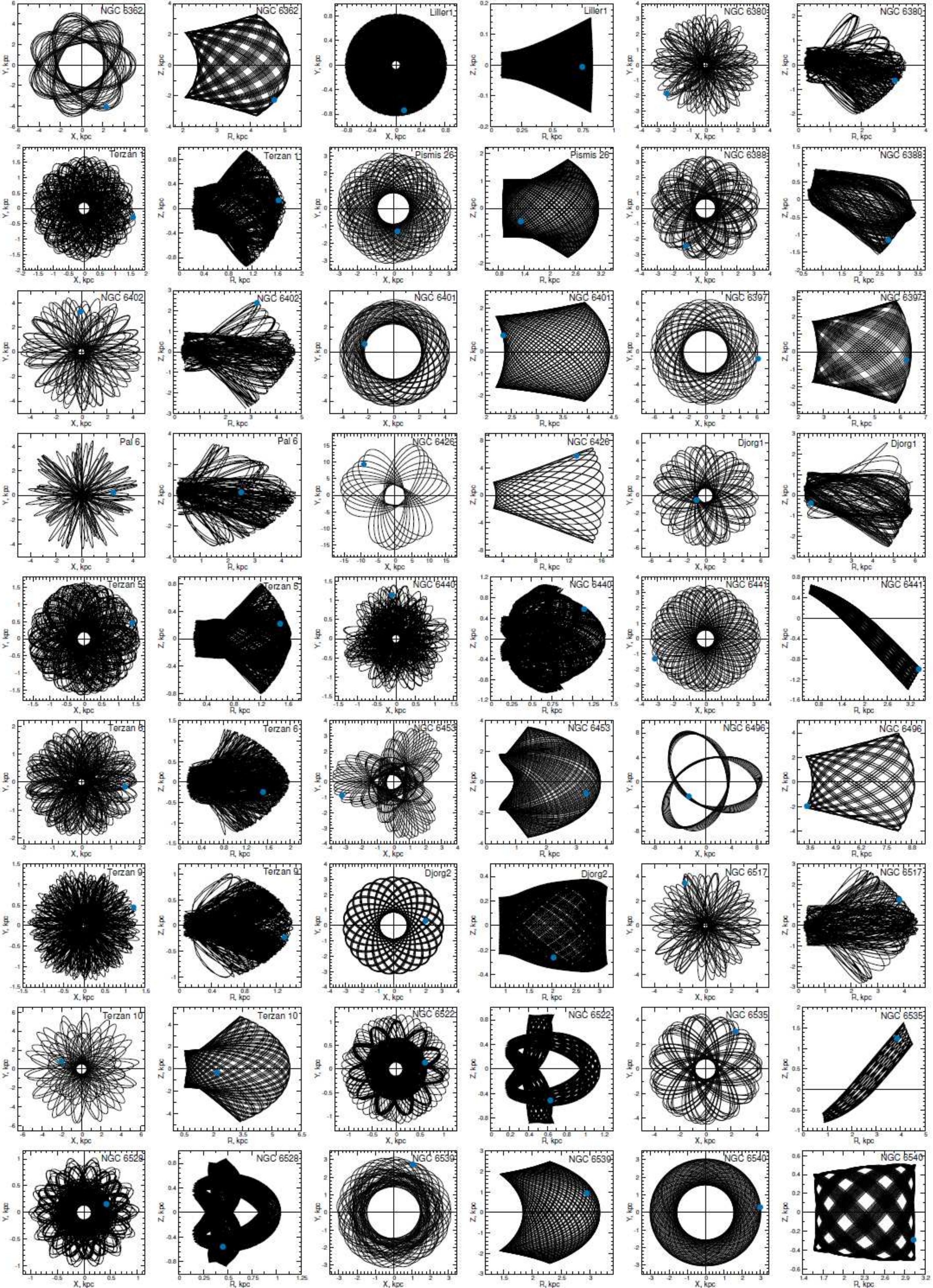}
 \centerline{Fig.~2 - continued}
\label{f4}
\end{center}}
\end{figure*}

\begin{figure*}
{\begin{center}
   \includegraphics[width=0.92\textwidth,angle=0]{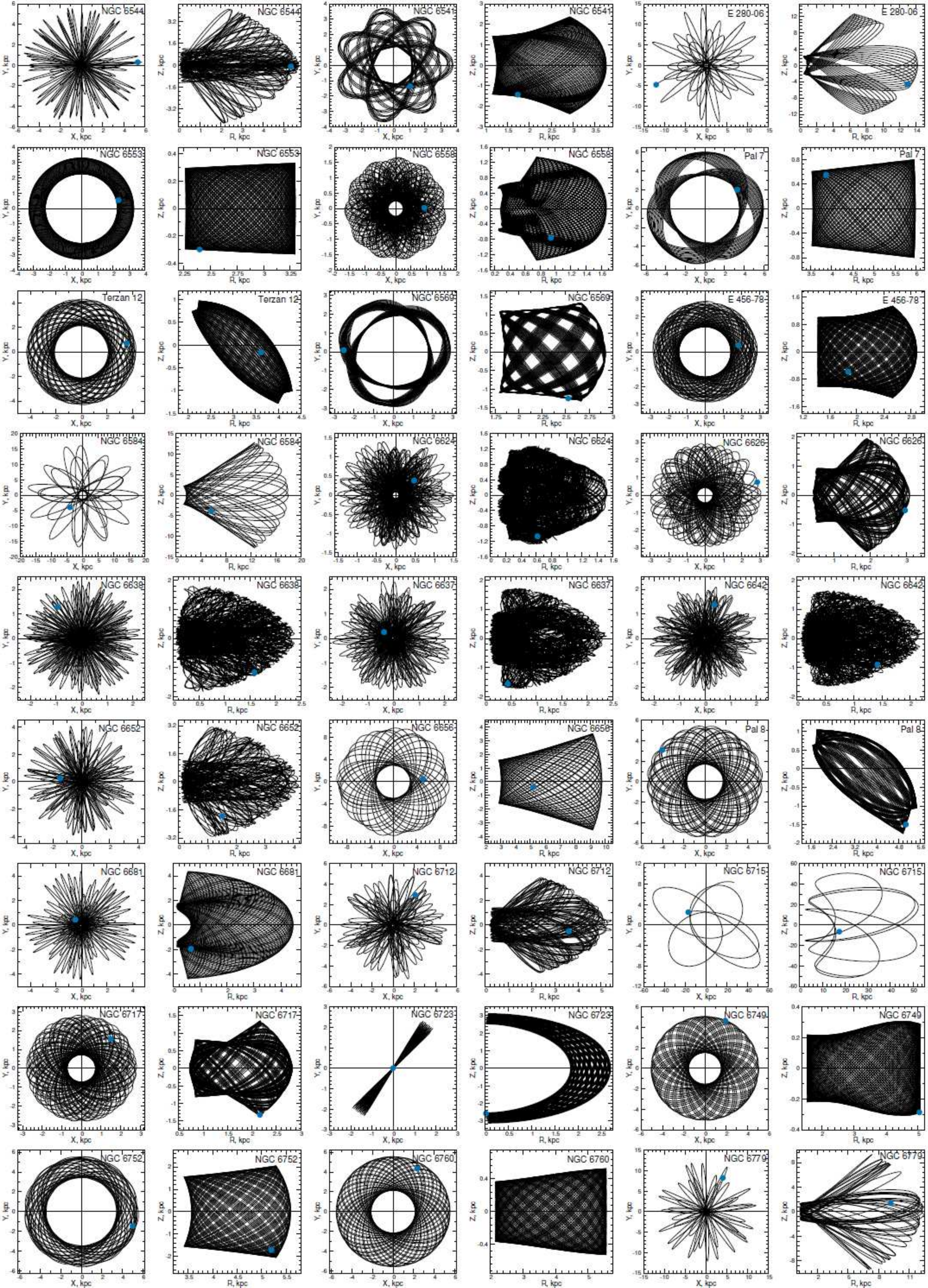}
 \centerline{{Fig.~2} - continued}
\label{f5}
\end{center}}
\end{figure*}

\begin{figure*}
{\begin{center}
   \includegraphics[width=0.92\textwidth,angle=0]{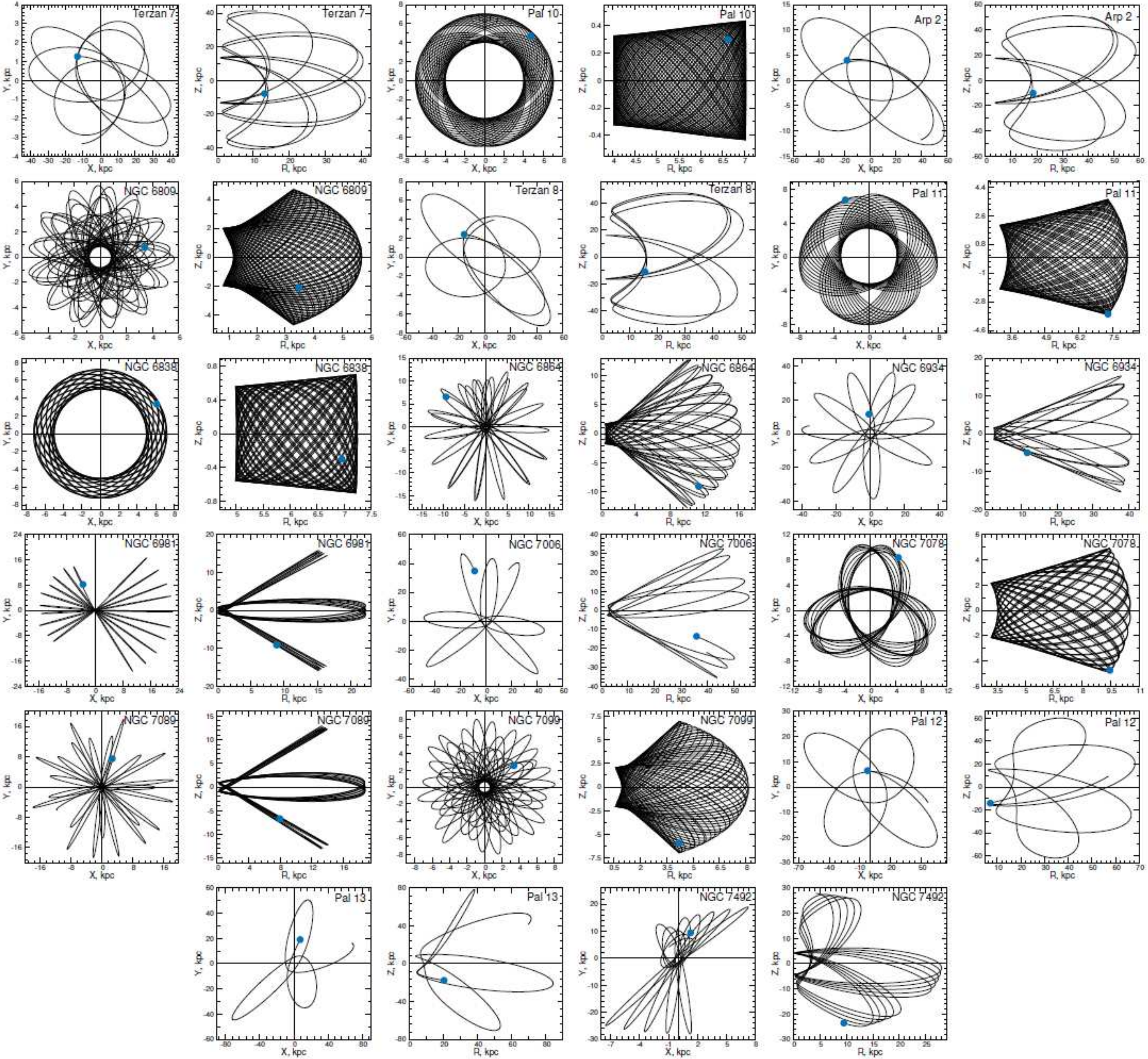}
 \centerline{Fig.~2 - continued}
\label{f6}
\end{center}}
\end{figure*}

\begin{figure*}
{\begin{center}
   \includegraphics[width=0.9\textwidth,angle=0]{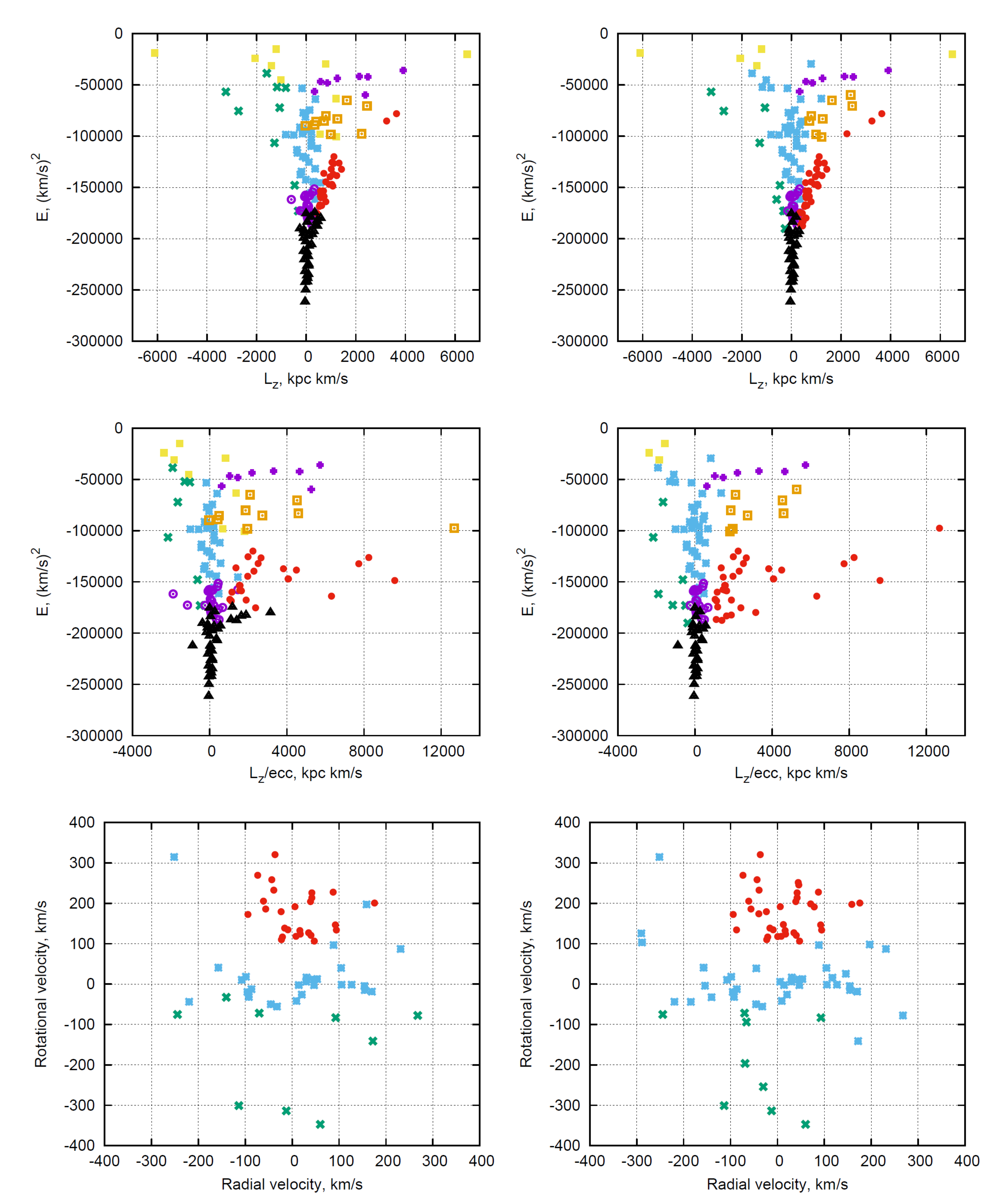}
\caption{The "$L_Z$ -- Energy" (top panel), "$L_Z/ecc$ -- Energy" (middle panel) and "Radial velocity -- Rotational velocity" (bottom panel) diagrams for GCs. Classification by Massari et al. (2019) (left-hand panels); the modified classification (right-hand panels). Colour-coded according to their belonging to different subsystems
(red symbols mark the disk, black is for the bulge, blue is for Gaia-Enceladus (Gaia-Sausage), orange for Helmi Streams, green for Sequoia, violet crosses for Sagittarius, yellow for the high-energy group, violet circles for the low-energy group. For visualisation purposes, clusters E 1 with
extremely negative $L_Z$ has not been plotted.  }
\label{fElz}
\end{center}}
\end{figure*}

\section{Conclusions}
For the first time since the emergence of the Gaia Data Release 2, a catalog of orbits for 152
galactic globular clusters, which form an almost complete known population, is presented.

The main orbital parameters of globular clusters are determined. The astrometric data catalog compiled by Vasiliev (2019) was used as data for calculating the initial 6d phase space (heliocentric positions and velocities of globular clusters) needed for orbit construction. The orbits were integrated for 5 Gyr backward.

For integrating the orbits, we used a recently obtained by Bajkova \& Bobylev (2016) the best-fit model of an axisymmetric Galactic potential with a dark halo in the form of NFW (Navarro et al., 1997) using data on circular velocities of Galactic objects in a wide range of galactocentric distances (data on HI region, masers, and catalog of Bhattacharjee et al. (2014)).

Based on the analysis of the obtained orbits and their properties, we have formed a modified composition of the subsystems of globular clusters, slightly different from the composition presented in Massari et al. (2019). This modification affected 27 GCs. The modified classification looks like a more correct one from the point of view of a greater similarity of the orbits of globular clusters included in their subsystem.

\section{Acknowledgements}
The authors would like to express gratitude to the anonymous referee for useful remarks, the consideration of which made it possible to significantly improve the article.
The authors are also grateful to Dr. Dambis A.K. for useful discussion.

\section{References}
\begin{itemize}

\item Bajkova, A.T. \& Bobylev, V.V. 2016, Ast. Lett., 42, 567 

\item Bajkova, A.T. \& Bobylev, V.V. 2017, OAst, 26, 72

\item Bajkova, A.T., Carraro, G., Korchagin, V.I., Budanova, N.O. \& Bobylev, V.V. 2020, ApJ, 895, 69

\item Baumgardt, H., Hilker, M., Sollima, A. $\&$ Bellini A. 2019, MNRAS, 482, 5138

\item Bhattacharjee, P., Chaudhury, S. $\&$ Kundu, S. 2014, ApJ, 785, 63

\item Bobylev, V.V. \& Bajkova A.T. 2016, Ast. Lett., 42, 1     

\item Harris, W. 2010, arXiv: 1012.3224

\item Gaia Collaboration, Helmi, A., van Leeuwen, F., McMillan, P.J.,
et al. 2018, A\&A, 616, 12

\item Irrgang, A., Wilcox, B., Tucker, E. \& Schiefelbein, L. 2013, A\&A, 549, 137

\item Koppelman, H.H. \& Helmi, A. 2020, arXiv: 2006.16283

\item Massari, D., Koppelman, H. H. \& Helmi, A. 2019, A\&A, 630, L4

\item Miyamoto, M. \& Nagai, R. 1975, PASJ, 27, 533

\item Myeong, G. C., Vasiliev, E., Iorio, G., Evans, N. W. \& Belokurov, V. 2019, MNRAS, 488, 1235

\item Navarro, J.F., Frenk, C.S. \& White, S.D.M. 1997, ApJ, 490, 493

\item Schonrich, R., Binney, J. \& Dehnen, W. 2010, MNRAS, 403, 1829

\item Vasiliev, E. 2019, MNRAS, 484, 2832

\item  Villanova, S., Monaco, L., Geisler D., et al. 2019, ApJ, 882, 174

\end{itemize}

\end{document}